\documentclass[ALICE,manyauthors]{cernphprep}

\usepackage[comma,square,numbers,sort&compress]{natbib}
\usepackage{hyperref}
\hypersetup{
    colorlinks=true,
    linkcolor=red,          
    citecolor=red,        
    filecolor=magenta,      
    urlcolor=black           
}
\usepackage{lineno}
\usepackage{version}
\usepackage{graphicx}
\usepackage{epstopdf}
\usepackage{epsfig}
\usepackage{amsmath, amssymb}
\usepackage{wasysym}
\usepackage{color}
\usepackage{xspace}
\usepackage{textcomp}
\usepackage{booktabs}
\usepackage{array}
\newcommand{\head}[2]{\multicolumn{1}{>{\centering\arraybackslash}p{#1}}{\textbf{#2}}}

\newcommand{\sqrts}{$\sqrt{s}$\xspace}
\newcommand{\pt}{$p_{\rm{T}}$}
\newcommand{\gevc}{\xspace GeV/$c$\xspace}
\newcommand{\gevcc}{\xspace GeV/$c^{2}$\xspace}
\newcommand{\mevc}{\xspace MeV/$c$\xspace}

\newcommand{\ptmean}{$\langle p_{\rm T} \rangle$}

\newcommand{\dndy}{d$N/$d$y$\xspace}
\newcommand{\dedx}{d$E/$d$x$\xspace}

\newcommand{\dcaxy}{DCA$_{xy}$\xspace}
\newcommand{\dcaz}{DCA$_z$\xspace}

\newcommand{\pbar}{$\rm\overline{p}$\xspace}
\newcommand{\tetackov}{$\langle\theta_{\rm{ckov}}\rangle$\xspace}
\graphicspath{{./figs/}}
\begin{document}%

\begin{titlepage}
\PHyear{2015}
\PHnumber{068}      
\PHdate{13 March}  
%

\title{Measurement of pion, kaon and proton production\\ in proton-proton collisions at $\sqrt{s}$ = 7 TeV}
\ShortTitle{Pion, kaon and proton production\\ in pp collisions at $\sqrt{s}$ = 7 TeV}   

\Collaboration{ALICE Collaboration\thanks{See Appendix~\ref{app:collab} for the list of collaboration members}}
\ShortAuthor{ALICE Collaboration} 


\begin{abstract}
The measurement of primary $\pi^{\pm}$, K$^{\pm}$, p and \pbar production at mid-rapidity ($|y| <$ 0.5) in proton-proton collisions at \sqrts = 7 TeV performed with ALICE (A Large Ion Collider Experiment) at the Large Hadron Collider (LHC) is reported. Particle identification is performed using the specific ionization energy loss and time-of-flight information, the ring-imaging Cherenkov technique and the kink-topology identification of weak decays of charged kaons. Transverse momentum spectra are measured from 0.1 up to 3\gevc for pions, from 0.2 up to 6\gevc for kaons and from 0.3 up to 6\gevc for protons. The measured spectra and particle ratios are compared with QCD-inspired models, tuned to reproduce also the earlier measurements performed at the LHC. Furthermore, the integrated particle yields and ratios as well as the average transverse momenta are compared with results at lower collision energies.
\end{abstract}
\end{titlepage}
\setcounter{page}{2}


\section{Introduction} 

The majority of the particles produced at mid-rapidity in proton-proton collisions are low-momentum hadrons not originating from the fragmentation of partons produced in scattering processes with large momentum transfer.
Their production, therefore, cannot be computed from first principles via perturbative Quantum Chromodynamics (pQCD). Currently available models describing hadron-hadron collisions at high energy, such as the event generators PYTHIA6~\cite{PYTHIA}, PYTHIA8~\cite{PYTHIA8,PYTHIA8i4Cx},  EPOS~\cite{EPOS,EPOS1} and PHOJET~\cite{PHOJET}, combine pQCD calculations for the description of hard processes with phenomenological models for the description of the soft component. The measurement of low-momentum particle production and species composition is therefore important as it provides crucial input for the modeling of the soft component and of the hadronization processes. 
Furthermore, it serves as a reference for the same measurement in Pb-Pb collisions to study the properties of the hot and dense strongly-interacting medium with partonic degrees of freedom, the Quark-Gluon Plasma (QGP), which is created in these collisions.
In this paper, the measurement of primary $\pi^{\pm}$, K$^{\pm}$, p and \pbar production at mid-rapidity in proton-proton collisions at \sqrts = 7 TeV using the  ALICE detector~\cite{PPR1,PPR2,ALICEperformance,JINST} is presented. Primary particles are defined as prompt particles produced in the collision including decay products, except those from weak decays of light flavour hadrons and muons. 
Pions, kaons and protons are identified over a wide momentum range by combining the information extracted from the specific ionization energy-loss (\dedx) measured in the Inner Tracking System (ITS) ~\cite{ITS_TDR} and in the Time Projection Chamber (TPC)~\cite{TPC_TDR}, the time of flight measured in the Time-Of-Flight (TOF) detector~\cite{TOF_TDR}, the Cherenkov radiation measured in the High-Momentum Particle Identification Detector (HMPID)~\cite{HMPID_TDR} and the kink-topology identification of the weak decays of charged kaons. 
Similar measurements in proton-proton collisions at \sqrts = 900 GeV and 2.76 TeV are reported in~\cite{900GeVspectra, 276ALICESpectra, CMS_spectra} and are included, together with lower energy data~\cite{PHENIX_2010, NA49r1, NA49r2, NA49r3, e735, E735prompt, UA5}, in the discussion of the evolution of particle production with collision energy.
Similar measurements at the LHC have also been performed in the forward region~\cite{LHCb}.

The paper is organized as follows. 
In Section 2 the ALICE experimental setup is described, focusing on the detectors and the corresponding Particle IDentification (PID) techniques relevant for the present measurement. Details on the event and track selection criteria and the corrections applied to the measured raw yields are also presented.
In Section 3 the results on the production of primary $\pi^{\pm}$, K$^{\pm}$ p and \pbar are shown. These include the transverse momentum (\pt \xspace) distributions and the \pt-integrated production yields of each particle species and the K/$\pi$ and p/$\pi$ ratios. The evolution with collision energy of the \pt-integrated particle yields, of their ratios and of their average transverse momenta \ptmean\xspace is also presented. In Section 4 particle spectra and their ratios (K/$\pi$ and p/$\pi$) are compared with models, in particular with different PYTHIA tunes~\cite{PYTHIA,PYTHIA8,PYTHIA8i4Cx,PYTHIA-P2011,PYTHIA-Z2}, EPOS~\cite{EPOS, EPOS1} and PHOJET~\cite{PHOJET}.
Section 5 concludes the paper summarizing the results.

\section{Experimental setup and data analysis}
\label{sec:analysis}

\subsection{The ALICE detector}
The ALICE detector was specifically optimized to reconstruct and identify particles over a wide momentum range thanks to the low material budget, the moderate magnetic field and the presence of detectors exploiting all the known PID techniques.
A comprehensive description of the ALICE experimental setup and performance can be found in~\cite{PPR1,PPR2,JINST,ALICEperformance}. In the following, the PID detectors relevant for the analysis presented in this paper are briefly described, namely ITS, TPC, TOF and HMPID. They are located in the ALICE central barrel in a $B = 0.5~\rm T$ solenoidal magnetic field directed along the beam axis. The ITS, TPC and TOF detectors cover the full azimuth ($\varphi$) and have a pseudorapidity coverage of $|\eta| < 0.9$, while the HMPID covers the pseudorapidity interval $|\eta| < 0.55$ and the azimuthal angle range $1.2^{\circ} < \varphi < 58.5^{\circ}$.\\
The ITS~\cite{ITS_TDR} is the innermost central barrel detector. It is composed of six cylindrical layers of silicon detectors, located at radial distances between 3.9 and 43 cm from the beam axis. The  two innermost layers are equipped with Silicon Pixel Detectors (SPD), the two intermediate ones are Silicon Drift Detectors (SDD), while the two outermost ones are Silicon Strip Detectors (SSD). 
The ITS provides high resolution tracking points close to the beam line, which allows us to reconstruct primary and secondary vertices with high precision, to measure with excellent resolution the Distance of Closest Approach (DCA) of a track to the primary vertex, and to improve the track \pt\xspace resolution. It is also used as a stand-alone tracker to reconstruct particles that do not reach the TPC or do not cross its sensitive areas. The SDD and SSD are equipped with analogue readout enabling PID via \dedx measurements with a relative resolution of about 10\%.\\
The TPC~\cite{TPC_TDR} is the main tracking detector of the ALICE central barrel. It is a large volume cylindrical chamber with high-granularity readout that surrounds the ITS covering the region 85 $< r <$ 247 cm and -250 $< z <$ +250  cm in the radial $r$ and longitudinal $z$ directions, respectively. It provides three-dimensional space points and specific ionization energy loss d$E$/d$x$ with up to 159 samples per track. The relative \dedx resolution is measured to be about 5.5\% for tracks that cross from the centre of the outer edge of the detector.\\
The TOF detector~\cite{TOF_TDR} is a large-area array of Multigap Resistive Plate Chambers (MRPC) with an intrinsic time resolution of 50 ps, including the electronic readout contribution. It is a cylindrical detector located at a radial distance 370 $< r <$ 399 cm from the beam axis. Particles are identified using simultaneously the time-of-flight information with the momentum and track length measured with the ITS and the TPC.\\
The HMPID~\cite{HMPID_TDR} is a single-arm proximity-focusing Ring Imaging CHerenkov (RICH) detector located at 475 cm from the beam axis.
The Cherenkov radiator is a 15-mm-thick layer of liquid C$_6$F$_{14}$ (perfluorohexane) with a refractive index of $n = 1.2989$ at a photon wave length $\lambda = 175$ nm, corresponding to a minimum particle velocity $\beta_{\rm{min}} = 0.77$. \\
In addition to the detectors described above that provide PID information, the VZERO system~\cite{VZERO} is used for trigger and event selection. It is composed of two scintillator arrays, which cover the pseudorapidity ranges $2.8<\eta<5.1$ and $-3.7<\eta<-1.7$. 

\subsection{Data sample, event and track selection}
\label{subsec:evtsel}
The results presented in this paper are obtained combining five independent analyses, namely ITS stand-alone, TPC-TOF, TOF, HMPID, kink, using different PID methods. The analyzed data are proton-proton collisions at 
\sqrts~=~7~TeV collected in 2010. During that period, the instantaneous luminosity at the ALICE interaction point was kept within the range 0.6-1.2$\times 10^{29}~\rm cm^{-2} s^{-1}$ to limit the collision pile-up probability. Only runs with a collision pile-up probability smaller than 4\% are used in this analysis, leading to an average pile-up rate of 2.5\%.
The number of events used in the five independent analyses is reported in Table \ref{tab:ranges_PID}.  
The data were collected using a minimum-bias trigger, which required a hit in the SPD or in at least one of the VZERO scintillator arrays in coincidence with the arrival of proton bunches from both directions. 
This trigger selection essentially corresponds to the requirement of having at least one charged particle in 8 units of pseudorapidity.
The contamination due to beam-induced background is removed off-line by using the timing information from the VZERO detector, which measures the event time with a resolution of about 1 ns, and the correlation between the number of clusters and track segments (tracklets) in the SPD~\cite{900GeVspectra}.\\
Selected events are further required to have a reconstructed primary vertex. For 87\% of the triggered events, the interaction vertex position is determined from the tracks reconstructed in TPC and ITS.
For events that do not have a vertex reconstructed from tracks, which are essentially collisions with low multiplicity of charged particles, the primary vertex is reconstructed from the SPD tracklets, which are track segments built from pairs of hits in the two innermost layers of the ITS. Overall, the fraction of events with reconstructed primary vertex, either from tracks or from SPD tracklets, is of 91\%.
Accepted events are required to have the reconstructed vertex position along the beam direction, $z$, within $\pm$ 10 cm from the centre of the ALICE central barrel. This ensures good rapidity coverage, uniformity of the particle reconstruction efficiency in ITS and TPC and reduction of the remaining beam-gas contamination.
In the following analyses two different sets of tracks are used: the global tracks, reconstructed using information from both ITS and TPC, and the ITS-sa tracks, reconstructed by using only the hits in the ITS. 
To limit the contamination due to secondary particles and tracks with wrongly associated hits and to ensure high tracking efficiency, tracks are selected according to the following criteria. The global tracks are required to cross over at least 70 TPC readout rows with a value of $\chi^{2} / N_{\rm{clusters}}$ of the
momentum fit in the TPC lower than 4, to have at least two clusters reconstructed in the ITS out of which at least one is in the SPD layers and to have a DCA to the interaction vertex in the longitudinal plane, \dcaz $<$ 2 cm. Furthermore, the daughter tracks of reconstructed kinks are rejected. This last cut is not applied in the kink analysis where a further \pt-dependent selection on the DCA of the selected tracks to the primary vertex in the transverse plane (\dcaxy) is requested. The global tracks that satisfy these selection criteria have a \pt\xspace resolution of 1\% at \pt\xspace= 1 \gevc and 2\% at \pt\xspace= 10 \gevc.
The ITS-sa tracks are required to have at least four ITS clusters out of which at least one in the SPD layers and three in the SSD and SDD, $\chi^{2} / N_{\rm{clusters}} < 2.5$ and a \dcaxy satisfying a \pt-dependent upper cut corresponding to 7 times the DCA resolution. The selected ITS-sa tracks have a maximum \pt\xspace resolution of 6\% for pions, 8\% for kaons and 10\% for protons in the \pt\xspace range used in the analysis.
Both global and ITS-sa tracks have similar resolution in the \dcaxy parameter that is 75 $\mu$m at \pt\xspace= 1\gevc and 20 $\mu$m at \pt\xspace = 15\gevc~\cite{DCA}, which is well reproduced in the simulation of the detector performance. 
The final spectra are calculated for $|y|<0.5$.

\subsection{Particle identification strategy}
\label{subsec:PIDstrategy}
To measure the production of $\pi^{\pm}$, K$^{\pm}$, p and \pbar over a wide~\pt\xspace range, results from five independent analyses, namely ITS-sa, TPC-TOF, TOF, HMPID and kink, are combined. Each analysis uses different PID signals in order to identify particles in the complementary \pt\xspace ranges reported in Table~\ref{tab:ranges_PID}. In the following, the PID strategies used by ITS-sa, TPC-TOF and TOF analyses are briefly summarized since they are already discussed in detail in~\cite{900GeVspectra, centralityPbPb},  while the HMPID analysis, presented here for the first time, and the kink analysis, modified with respect to that described in~\cite{900GeVspectra}, are presented in more detail.

\begin{table}[tp]
\vspace{0.5cm}
  \centering
  \begin{tabular}{c|cccc}
  \hline
    Analysis & \# of events &  $\pi$ & K & p \\
    \hline
    \hline
    ITS-sa   & $5.4\cdot 10^{7}$&  0.1-0.6  & 0.2-0.5  & 0.3-0.6   \\
    TPC--TOF & $5.4\cdot 10^{7}$& 0.25-1.2 & 0.3-1.2  & 0.45-2.0  \\
    TOF      & $5.4\cdot 10^{7}$ & 0.5-2.5  & 0.5-2.4  & 0.8-4.0  \\
    HMPID   & $8.1\cdot 10^{7}$ & 1.5-3.0  & 1.5-3.0  & 1.5-6.0  \\
    Kink    & $16.9\cdot 10^{7}$ &    -      & 0.2-6.0   &   -  \\
    \hline
 \end{tabular}
 \caption{Number of analyzed events and \pt\xspace range (GeV/$c$) covered by each analysis.}
 \label{tab:ranges_PID}
\end{table}

\subsubsection{ITS stand-alone analysis}\label{sec:itssa}
In this analysis ITS-sa tracks are used and particles are identified by comparing the \dedx measurement provided by the ITS detector with the expected values at a given momentum $p$ under the corresponding mass hypotheses.  
\begin{figure}
\centering
\includegraphics[width=8.5cm]{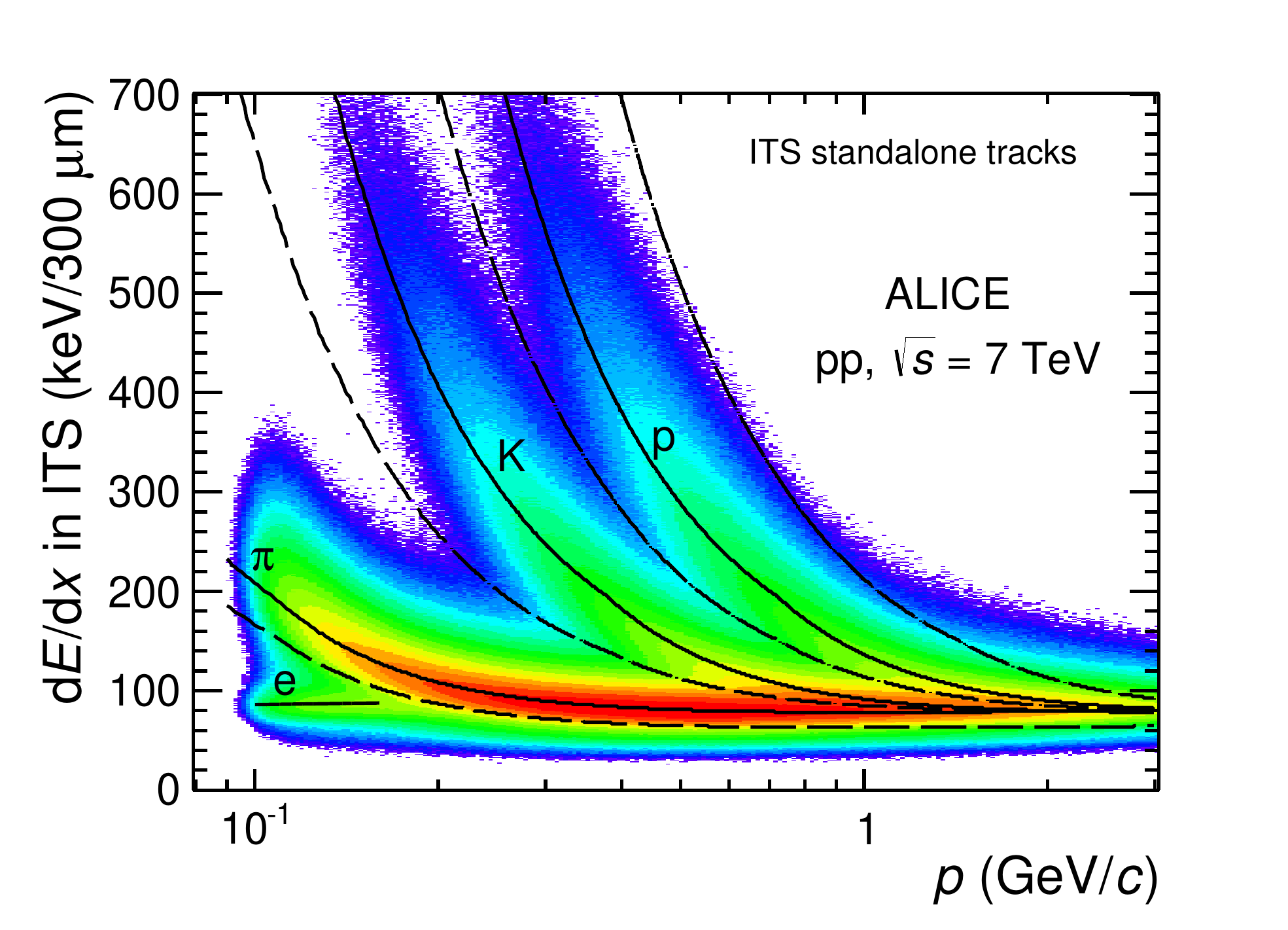}
\caption{Distribution of \dedx as a function of momentum ($p$) measured in the ITS using ITS-sa tracks in $|\eta|<0.9$. The continuous curves represent the parametrization of \dedx for e, $\pi$, K and p while the dashed curves are the bands used in the PID procedure.
}
\label{fig:ITSsa:dEdx}
\end{figure}
In Fig.~\ref{fig:ITSsa:dEdx}, the measured \dedx values are shown as a function of track momentum together with the curves of the energy loss for the different particle species, which are calculated
using the PHOBOS parametrization~\cite{Back2007} of the Bethe-Bloch curves at large $\beta\gamma$ and with a polynomial to correct for instrumental effects. A single identity is assigned to each track according to the mass hypothesis for which the expected specific energy-loss value is the closest to the measured \dedx for a track with momentum $p$.
 No explicit selection on the difference between the measured and expected values is applied except for a lower limit on pions set to two times the \dedx resolution ($\sigma$) and an upper limit on protons given by the mid-point between the proton and the deuteron expected \dedx.
The ITS \dedx is calculated as a truncated mean of three or four \dedx values provided by the SDD and SSD layers. The truncated mean is the average of the lowest two \dedx values in case signals in all the four layers are available, or as a weighted average of the lowest (weight 1) and the second lowest (weight 1/2) values in the case where only three \dedx samples are measured. 
Even with this truncated mean approach, used to reduce the effect of the tail of the Landau distribution at large \dedx, the small number of samples results in residual non-Gaussian tails in the \dedx distribution, which are partially reproduced in simulation. 
These non-Gaussian tails increase the misidentification rate, e.g. pions falling in the kaon identification bands. The misidentification probability is estimated using a Monte-Carlo simulations where the particle abundances were adjusted to those observed in data.
This correction is at most 10\% in the \pt\xspace range of this analysis.
In order to check possible systematic effects due to these non-Gaussian tails and their imperfect description in Monte-Carlo simulations, the analysis was repeated with different strategies for the particle identification, namely using a 3$\sigma$ compatibility band around the  expected \dedx curves and extracting the yields of pions, kaons and protons using the unfolding method described in ~\cite{900GeVspectra}, which is based on fits to the \dedx distributions in each \pt\xspace interval. The difference among the results from these different analysis strategies is assigned as a systematic uncertainty due to the PID.

\subsubsection{TPC--TOF analysis}
\begin{figure}
\centering
\includegraphics[width=8.5cm]{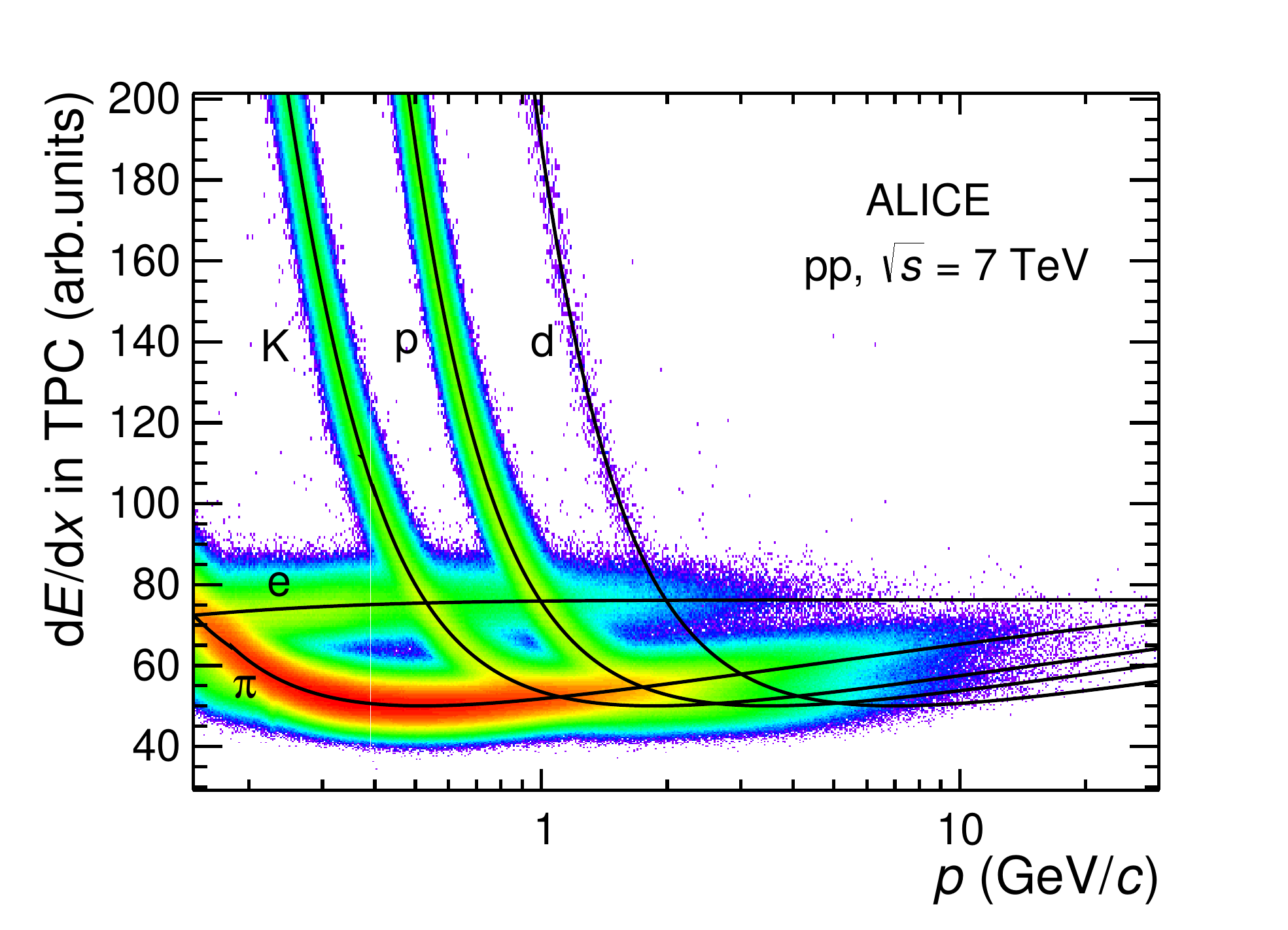}
\caption{Distribution of d$E$/d$x$ as a function of momentum ($p$) measured in the TPC using global tracks for $|\eta| < 0.9$. The continuous curves represent the Bethe-Bloch parametrization.
}
\label{fig:TPC:dEdx}
\end{figure}

\begin{figure}
\centering
\includegraphics[width=8.5cm]{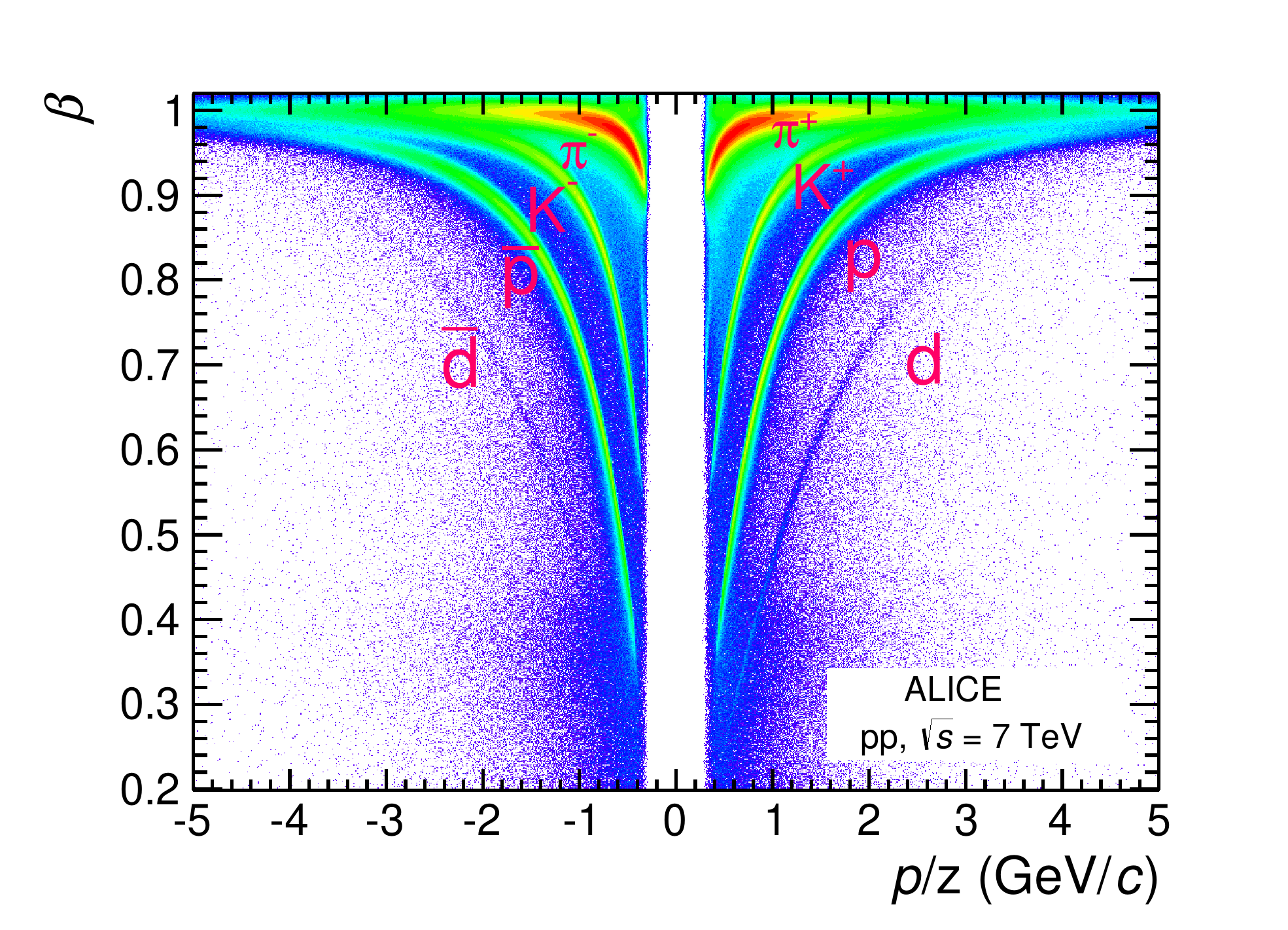}
\caption{Particle velocity $\beta$ measured by the TOF detector as a function of the rigidity $p/z$, where $z$ is the particle charge, for $|\eta| < 0.9$.}
\label{fig:TOF:beta}
\end{figure}

In this analysis global tracks are used and particle identification is performed by comparing the measured PID signals in the TPC and TOF detectors (\dedx, time of flight) with the expected values for different mass hypotheses.  
An identity is assigned to a track if the measured signal differs from the expected value by less than three times its resolution $\sigma$. 
For pions and protons with \pt~$<$~0.6~\gevc and kaons with \pt~$<$~0.5~\gevc, a compatibility within 3$\sigma$ is required on the \dedx  measurement provided by the TPC computed as a truncated mean of the lowest 60\% of the  available \dedx samples. The \dedx resulting from this truncated mean approach is Gaussian and it is shown in Fig.~\ref{fig:TPC:dEdx} as a function of the track momentum together with the expected energy-loss curves (see ~\cite{eloss} for a discussion about the \dedx parametrization). 
Above these \pt\xspace thresholds, i.e. \pt\xspace $\geq$ 0.6 \gevc for pions and protons and \pt\xspace $\geq$ 0.5 \gevc for kaons, a three $\sigma$ requirement is applied to both the \dedx  measurement provided by the TPC and the time of flight $t_{\rm{tof}}$ provided by the TOF detector. The time of flight $t_{\rm{tof}}$, as will be described in more detail in the next section, is the difference between the arrival time $\tau_{\rm TOF}$ measured with the TOF detector and the event start time $t_0$, namely $t_{\rm{tof}}=\tau_{\rm{TOF}}-t_{0}$. The additional condition on the TOF signal helps in extending the particle identification on a track-by-track basis to higher \pt\xspace where the TPC separation power decreases. 
The particles for which the time-of-flight signal is available are a sub-sample of the global tracks reconstructed using ITS and TPC information.
The time-of-flight information is not available for tracks that cross inactive regions of the TOF detector, for particles that decay or interact with the material before the TOF and for tracks whose trajectory, after prolongation from the TPC outer radius, is not matched with a hit in the TOF detector. The fraction of global tracks with associated time-of-flight information (TOF matching efficiency) depends on the particle species and  \pt\xspace as well as on the fraction of the TOF active readout channels. For the data analysis presented in this paper the matching efficiency
increases with increasing \pt\xspace until it saturates, e.g. at about 65\% for pions with \pt\xspace $>$ 1\gevc.
In Fig.~\ref{fig:TOF:beta} the velocity $\beta$ of the tracks, computed from the trajectory length measured with the ITS and TPC and the time of flight measured  with the TOF, is reported as a function of the rigidity $p/z$, where $z$ is the charge assigned based on the measured direction of the track curvature.
More than one identity can be assigned to a track if
it fulfils PID and rapidity selection criteria for different particle species.
The frequency of such cases is at most 0.5\% in the momentum range used in this analysis.
The misidentification of primary particles is computed and corrected for using Monte-Carlo simulations. It is at most 2\% for pions and protons and 8\% for kaons in the considered \pt\xspace ranges. The correction of the raw spectra for the misidentified particles provides also a way to remove the overestimation of the total number of particles introduced by the possibility, described above, to assign more than one identity to a track.

\subsubsection{TOF analysis}
\label{sec:tof-analysis}
\begin{figure}
\centering
\includegraphics[width=8.5cm]{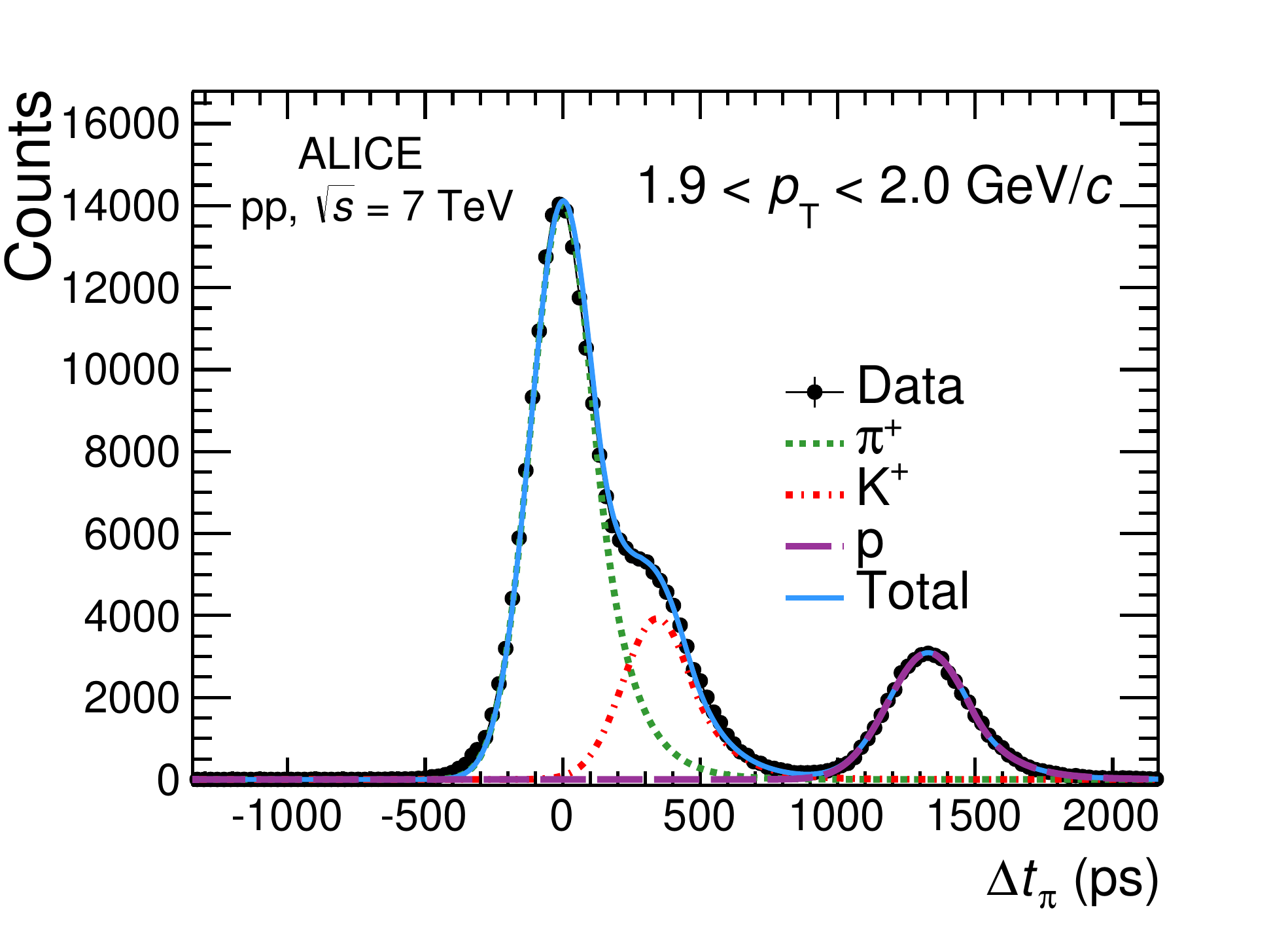}
\caption{Distribution of $\Delta t_{i}$ assuming the pion mass hypothesis in the transverse momentum interval 1.9~$<$~\pt~$<$~2.0~\gevc. The data (black points) are fitted with a function (light blue line) that is the sum of the signal due to pions (green dotted line) and the two populations corresponding to kaons (red dotted line) and protons (purple dashed line).}
\label{fig:TOF_fit}
\end{figure}

This analysis uses the sub-sample of global tracks for which a time-of-flight measurement is available.
The PID procedure utilizes a statistical unfolding approach that provides a \pt\xspace reach higher than the three $\sigma$ approach described in the previous section. The procedure is based on the comparison between the measured time of flight from the primary vertex to the TOF detector, $t_{\rm{tof}}$, and the time expected under a given mass hypothesis, $t^{\rm{exp}}_{i}$ ($i$ = $\pi$, K, p), namely on the variable $\Delta t_{i} = t_{\rm{tof}} - t^{\rm{exp}}_{i}$. 
As mentioned in the previous section, the time of flight $t_{\rm{tof}}$ is defined as the difference between the time measured with the TOF detector $\tau_{\rm TOF}$ and the event start time $t_0$. The $t_{0}$ value is computed from the analyzed tracks themselves on an event-by-event basis, using a combinatorial algorithm which compares the measured $\tau_{\rm{TOF}}$ with the expected ones for different mass hypotheses. The track under study is excluded to avoid any bias in the PID procedure~\cite{900GeVspectra, TOF_TDR}. In case the TOF $t_{0}$ algorithm fails, the average beam-beam interaction time is used.
The former approach provides a better $t_{0}$ resolution, but requires at least three reconstructed tracks with an associated TOF timing measurement. 
The yield of particles of species $i$ in a given \pt\xspace interval is obtained by fitting the distribution of the variable $\Delta t_{i}$ obtained from all the tracks regardless of the method used to compute the $t_{0}$. This distribution is composed of the signal from particles of species $i$, which is centered at  $\Delta t_{i}=0$, and two distinct populations corresponding to the other two hadron species, $j,k \neq i$. The $\Delta t_{i}$ distribution is therefore fitted with the sum of three functions $f(\Delta t_i)$, one for the signal and two for the other hadron species, as shown in Fig.~\ref{fig:TOF_fit}. The 
$f(\Delta t_i)$ functional forms are defined using the data in the region of clear species separation. The TOF signal is not purely Gaussian and it is described by a function $f(\Delta t_i)$  that is composed of a Gaussian term and an exponential tail at high $\Delta t_{i}$ mainly due to tracks inducing signals in more than one elementary detector readout element~\cite{TOF_TDR}. 
The raw yield of the species $i$ is given by the integral of the signal fit function.
The reach in \pt\xspace of this PID method depends on the resolution of $\Delta t_{i}$ that is the combination of the TOF detector intrinsic resolution, the uncertainty on the start time and the tracking and momentum resolution. Its value, for the data used in this analysis, is about 120 ps leading to 2$\sigma$ pion-kaon and kaon-proton separation at \pt\xspace = 2.5\gevc and \pt\xspace = 4.0\gevc, respectively.
This PID procedure has the advantage of not requiring a Monte-Carlo based correction for misidentification because the contamination under the signal of particles of species $i$ due to other particle species is accounted for by the background fit functions.

\subsubsection{HMPID analysis}
\begin{figure}
\includegraphics[width=8.5cm]{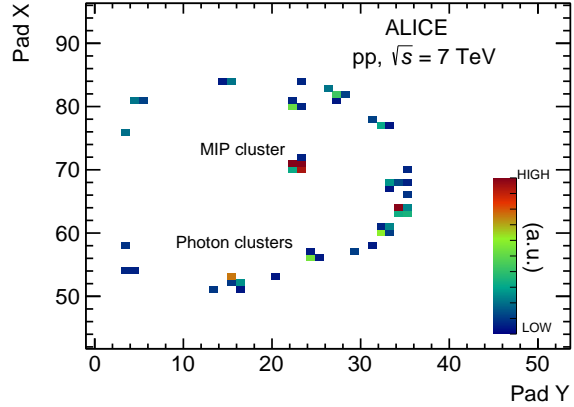}
\centering
\caption{Display of a Cherenkov ring detected in a module of HMPID for an inclined track crossing the detector. 
The colors are proportional to the pad charge signal. 
}
\label{fig:HMPID:Ring}
\end{figure}

The HMPID is a Ring Imaging Cherenkov (RICH) detector in a proximity focusing layout in which the primary ionizing charged particle generates Cherenkov light inside a liquid C$_6$F$_{14}$ radiator~\cite{HMPID_TDR}. The UV photons are converted into photoelectrons in a thin CsI film of the PhotoCathodes (PCs) and the photoelectrons are amplified in an avalanche process inside a Multi-Wire Proportional Chamber (MWPC) operated with CH$_4$. To obtain position sensitivity for the reconstruction of the Cherenkov rings, the PCs are segmented into pads. The final image of a Cherenkov ring is then formed by a cluster of pads (called ``MIP'' cluster) associated to the primary ionization of the particle and the photoelectron clusters associated to Cherenkov photons. In Fig.~\ref{fig:HMPID:Ring} a typical Cherenkov ring is shown.

\begin{figure}
\centering
\includegraphics[width=8.5cm]{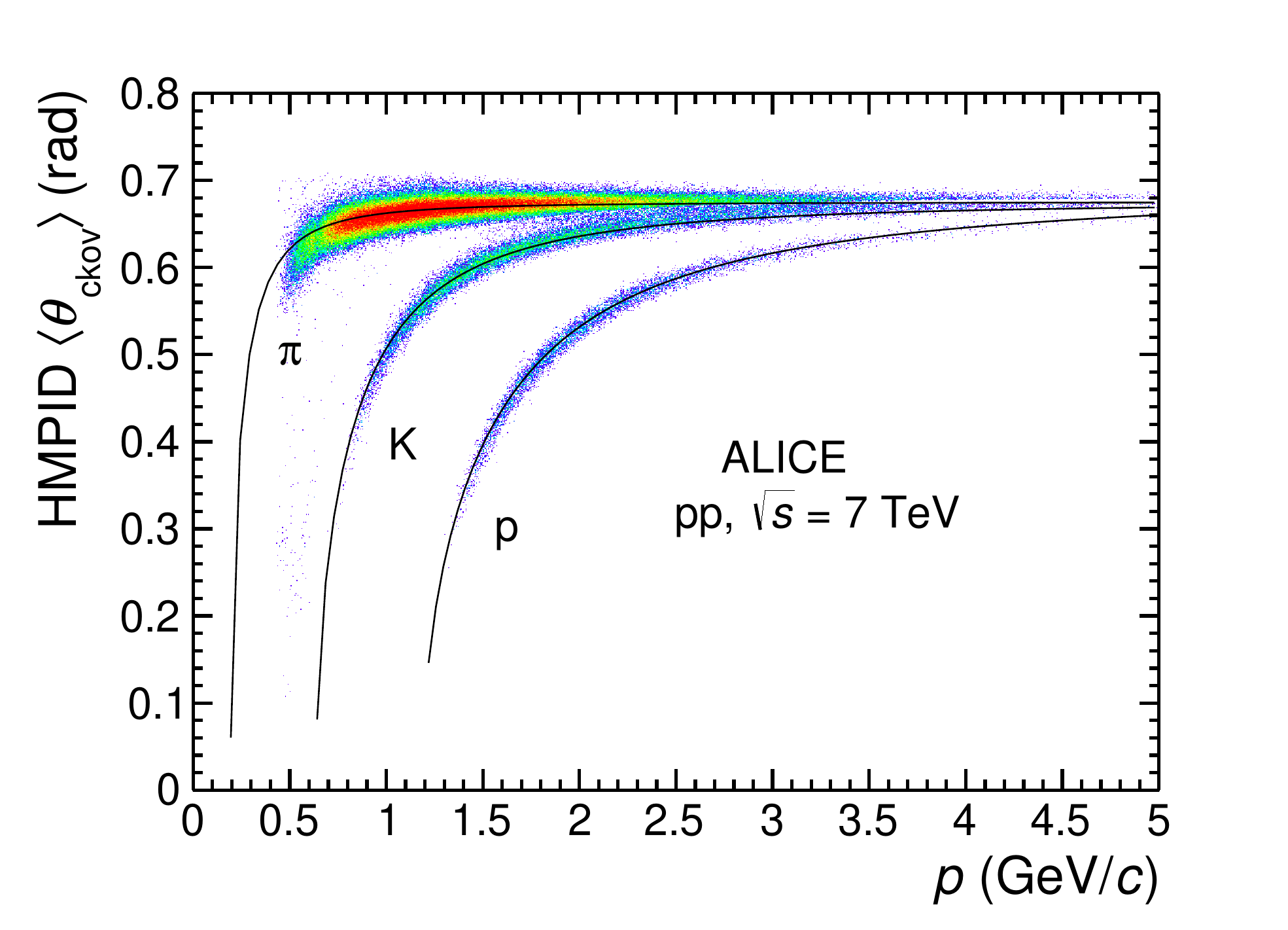}
\caption{Mean Cherenkov angle \tetackov measured with HMPID in its full geometrical acceptance as a function of the particle momentum $p$ for positively and negatively charged
tracks. The solid lines represent the theoretical curves for each particle species.} 
\label{fig:HMPID:Cherenkov}
\end{figure}

In this analysis, the sub-sample of global tracks that reach the HMPID detector and produce the Cherenkov rings is used. 
Starting from the photoelectron cluster coordinates on the photocathode, a back-tracking algorithm calculates the corresponding single photon Cherenkov angle by using the impact angle of a track extrapolated from the central tracking detectors up to the radiator volume. 
A selection on the distance ($d_{\rm{MIP-trk}}$) computed on the cathode plane between the centroid of the MIP cluster and the track extrapolation, set to $d_{\rm{MIP-trk}}$~$<$~5~cm, rejects fake associations in the detector.
Background discrimination is performed using the Hough Transform Method~(HTM)~\cite{DiBari2003}.\\
The mean Cherenkov angle \tetackov is obtained if at least three 
photoelectron clusters 
are detected. 
For a given track,  \tetackov is computed as the weighted average of the single photon angles (if any) selected by HTM. 
Pions, kaons and protons become indistinguishable at high momentum when the resolution on \tetackov reaches 3.5 mead. 
The angle \tetackov as a function of the track momentum is shown in Fig.~\ref{fig:HMPID:Cherenkov}, where the solid lines represent the $\theta_{\rm{ckov}}$ dependence on the particle momentum
\begin{equation}
 \theta_{\rm{ckov}} = \cos^{-1} \frac{\sqrt{p^2+m^2}}{np},
\end{equation}
where $n$ is the refractive index of the liquid radiator, $m$ the mass of the particle and $p$ its momentum.
\\
This analysis is performed for $p$~$>$1.5~GeV/c, where pions, kaons and protons produce a ring with enough photoelectron clusters to be reconstructed. If the track momentum is below the threshold to produce Cherenkov photons, background clusters could be wrongly associated to the track. As an example  the few entries visible in Fig.~\ref{fig:HMPID:Cherenkov} between the pion and kaon bands at low $\langle\theta_{\rm{ckov}}\rangle$ correspond to wrong associations of clusters with a kaon or a proton below the threshold to produce Cherenkov photons.

\begin{figure}
\centering
\includegraphics[width=8.5cm]{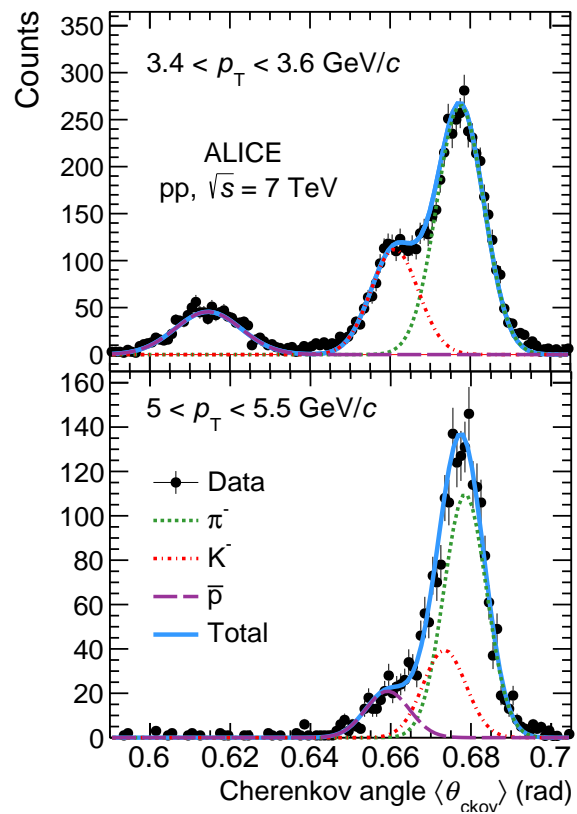}
\caption{Distributions of \tetackov measured with the HMPID in the two narrow \pt\xspace intervals 3.4~$<$~\pt~$<$~3.6~GeV/$c$ (top) and 5~$<$~\pt~$<$~5.5~GeV/$c$ (bottom) for tracks from negatively-charged particles. Solid lines represent the total fit (sum of three Gaussian functions). Dotted lines correspond to pion, kaon and proton signals. The background is negligible. 
}
\label{fig:FIT_HMPID}
\end{figure}

\begin{figure}
\centering
\includegraphics[width=8.5cm]{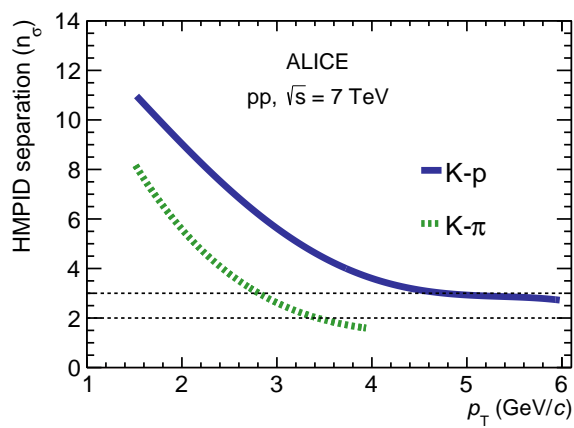}
\caption{Separation power ($n_{\sigma}$) of hadron identification in the HMPID as a function of \pt. The separation n$_{\sigma}$ of pions and kaons (kaons and protons) is defined as the difference between the average of the Gaussian distributions of \tetackov for the two hadron species divided by the average of the Gaussian widths of the two distributions.}
\label{fig:HMPID:Sep}
\end{figure}

The particle yields are extracted from a fit to the Cherenkov angle distribution in narrow transverse momentum intervals. In Fig.~\ref{fig:FIT_HMPID}, examples of the reconstructed Cherenkov angle distributions in two narrow \pt~intervals (3.4~$<$~\pt~$<$~3.6~GeV/$c$ and 
5~$<$~\pt~$<$~5.5~GeV/$c$) for negatively-charged tracks are shown. \\
The background, mainly due to noisy pads and photoelectron clusters from other rings overlapping to the reconstructed one, is negligible in the momentum range considered in this analysis.
The fit function (shown as a solid line in Fig.~\ref{fig:FIT_HMPID}) is a sum of three Gaussian functions, one for each particle species (dashed lines), whose mean and sigma are fixed to the Monte-Carlo values.   
The extracted separation power of hadron identification in the HMPID as a function of \pt~is shown in Fig.~\ref{fig:HMPID:Sep}. The separation between pions and kaons (kaons and protons) is expressed as the difference between the means of the \tetackov angle Gaussian distributions for the two given particle species ($\Delta_{\rm{\pi,K}}$ or $\Delta_{\rm{K,p}}$) divided by the average of the Gaussian widths of the two distributions, i.e. ($\sigma_{\rm\pi}$+$\sigma_{\rm K}$)/2 or ($\sigma_{\rm K}$+$\sigma_{\rm p}$)/2.
A separation at 3$\sigma$ level in \tetackov is achieved up to \pt\xspace= 3\gevc for K-$\pi$ and up to \pt\xspace= 5 \gevc  for K-p. The separation at 6 \gevc for K-p can be extrapolated from the curve and it is about 2.5$\sigma$.
The HMPID geometrical acceptance is about 5\% for tracks with high momentum. Therefore the analysis of HMPID required to analyze a larger data sample with respect to the other PID methods, as reported in Table \ref{tab:ranges_PID}.
The total efficiency is the convolution of the tracking, matching and PID efficiencies.
The PID efficiency of this method is determined by the Cherenkov angle reconstruction efficiency. It has been computed by means of Monte-Carlo simulations and it reaches 90\% for particles with velocity $\beta\sim$~1. As a cross check, the PID efficiency has been determined using clean samples of protons and pions from $\Lambda$ and K$^0_{\rm s}$ decays. The measured efficiency agrees within the statistical uncertainties with the Monte-Carlo estimates, in the momentum range 1.5 $<$ \pt\xspace$<$ 6 \gevc. Moreover, the correction due to the $d_{\rm{MIP-trk}}$ cut is computed from the same sample of identified protons and pions from $\Lambda$ and K$^{0}_{\rm s}$ decays.

\subsubsection{Kink analysis} \label{kink_analysis}
Charged kaons can also be identified in the TPC by reconstructing their 
weak-decay vertices, which exhibit a characteristic kink topology defined by a decay vertex with two tracks (mother and daughter) having the same charge. 
This procedure extends the measurement of charged
kaons on a track-by-track basis to \pt\xspace= 6\gevc.
The  algorithm for the kink reconstruction is
applied inside a fiducial volume of the TPC, namely 130~$ < R < $~200 cm, needed to reconstruct  both the mother and the daughter tracks. The mother track is selected with similar
criteria as the global tracks (Section 2.2), but
with a looser selection on the minimum number of TPC clusters, which is set to 20, and a wider rapidity range set to $|y|< 0.7$  to increase the statistics of kink candidates. No selections are applied on the charged daughter track.
The reconstructed invariant mass $M_{\mu\nu}$ is calculated  assuming the charged daughter track to be a muon and the undetected neutral daughter track to be a neutrino. The neutrino momentum is the difference between the measured momenta of the mother particle and of the charged daughter.
\begin{figure}
\centering
\includegraphics[width=8.5cm]{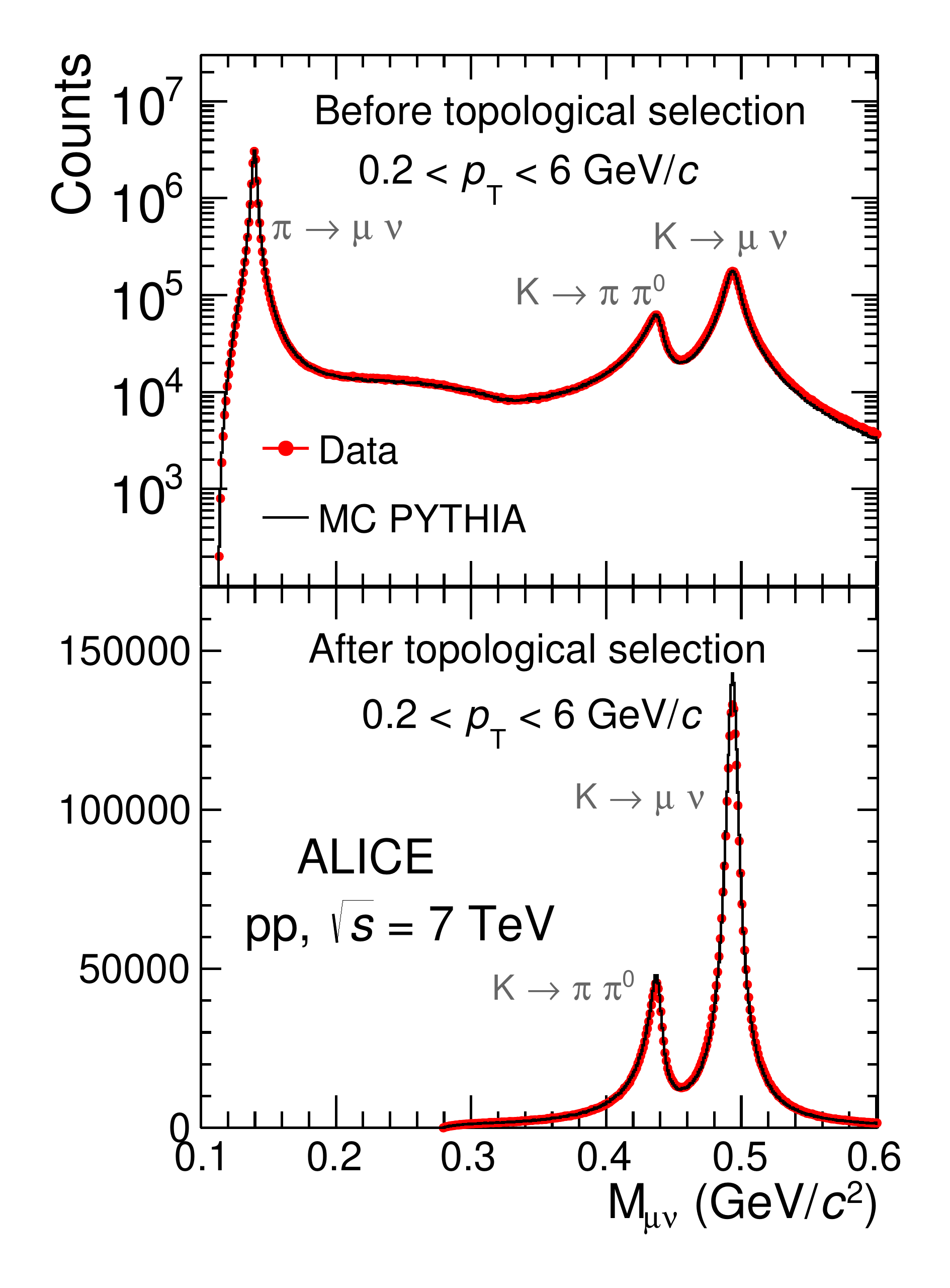}
\caption{Kink invariant mass $M_{\mu\nu}$ in data (red circles) and Monte-Carlo (black line) for summed particles and antiparticles, integrated over the mother transverse momentum range 0.2  $<$ \pt\xspace $< 6.0 $\gevc and $|y| < 0.7$ before (top panel) and after (bottom panel) the topological selections, based mainly on the $q_\mathrm{T}$ and the maximum decay opening angle.
}
\label{fig:Kinks:mass}
\end{figure}
The $M_{\mu\nu}$ distribution, for summed positive and negative charges, integrated over the mother transverse momentum range 0.2 $<$ \pt\xspace $< 6.0 $\gevc  is reported in the top panel of Fig.~\ref{fig:Kinks:mass} for both data and PYTHIA simulations normalized to the same number of entries. Three peaks are present: one centered on the kaon mass due to the kaon decays K$\rightarrow\mu + \nu_{\mu}$ (branching ratio BR = 63.55\%), one centered at  $M_{\mu\nu}$ = 0.43\gevcc due to the {K$\rightarrow\pi + \pi^0$} decay (BR = 20.66\%), whose kinematics is calculated with wrong mass assumptions, and the peak due to pion decays $\pi\rightarrow\mu + \nu_{\mu}$ (BR = 99.99\%). The width of the peaks reflects the momentum resolution of the detector, which is well reproduced in Monte-Carlo simulations.
The two-body kinematics of the kink topology allows one to separate kaon decays from the main source of background due to charged pion decays~\cite{900GeVspectra}.
In the  $\mu +\nu_{\mu}$ channel, the upper limit of the $q_\mathrm{T}$ variable, where $q_\mathrm{T}$ is defined as the transverse momentum of the daughter track with respect to the mother's direction, is 236~MeV/$c$ for muons from kaon decays and 30 MeV/$c$ for muons from pion decays. 
To remove most of the pion decays,  a $q_\mathrm{T}>$~120~MeV/$c$ selection is applied.
The background is further reduced by rejecting kink decays for which the decay angle, namely  the angle between the momenta of the mother and the charged daughter tracks is larger than the maximum angle allowed in the hypothesis K$\rightarrow\mu + \nu_{\mu}$.
The bottom panel of Fig.~\ref{fig:Kinks:mass}  shows the invariant mass distribution of the kaon candidates with mother transverse  momentum 0.2 $<$ \pt\xspace$ < 6.0$\gevc after the topological selection criteria for kaon identification (mainly the $q_\mathrm{T}$  and decay angle cuts) are applied. It is evident that only the two peaks coming from kaon decays are present while the pion background peak is removed. 
The broad structure on the left originates from the three-body decays of kaons. 
The agreement between data and simulations in this figure (Fig.~\ref{fig:Kinks:mass}) is better than 8\%.
Most of the selected mother tracks have a \dedx in the TPC which is compatible with the values expected for kaons.
Tracks outside 3.5$\sigma$ from the expected kaon \dedx have been removed to attain a purity $>$~97~\% in the \pt\xspace range studied in this analysis. These rejected tracks are less than 4\%, have \pt\xspace$ < $ 0.8\gevc and are, according to Monte-Carlo studies, pions. 
The raw kaon spectra are obtained from the integral of the invariant mass distribution computed in narrow \pt\xspace intervals after the topological selection criteria on  the $q_{T}$,  the decay opening angle, and the compatibility with the expected \dedx for kaons are applied.
The kaon misidentification is computed and corrected for by using Monte-Carlo simulations. It depends on the mother's transverse momentum  with a maximum value of 3.6\% at 0.8\gevc and a minimum of 2\% at 1\gevc, remaining almost flat up to \pt\xspace $=$ 6\gevc. Its average value in the \pt\xspace range considered in this analysis is 2.1\%.

\subsection{Correction of raw spectra}
To obtain the \pt\xspace distributions of primary $\pi$, K and p, the contribution of secondaries is subtracted from the raw spectra. Then, the spectra are corrected for the PID efficiency, the misidentification probability, the acceptance, the reconstruction and the selection efficiencies according to 

\begin{equation}\label{eq:corr}
\frac{{\rm d}^{2} N}{{\rm d} p_{\rm{T}} {\rm d} y} =N_{\rm raw}(p_{\rm{T}})\frac{1}{\Delta p_{\rm{T}}\Delta y} \frac{1-s(p_{\rm{T}})}{\varepsilon(p_{\rm{T}})}\cdot f(p_{\rm{T}}), 
\end{equation}
where $N_{\rm raw}(p_{\rm{T}})\frac{1}{\Delta p_{\rm{T}}\Delta y}$ is the raw yield in a given \pt\xspace interval, $s(p_{\rm{T}})$ is the total contamination including effects of secondary and misidentified particles, $\varepsilon(p_{\rm{T}})$ is the acceptance$\times$efficiency including PID efficiency, detector acceptance, reconstruction and selection efficiencies and $f(p_{\rm{T}})$ is an additional factor to correct for imperfections of the cross sections for antiparticle interactions with the material used in the GEANT3 code. 
 
\begin{figure}
\centering
\includegraphics[width=8.5cm]{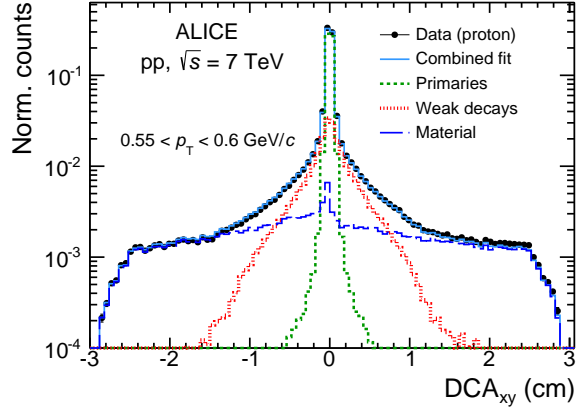}
\caption{Proton \dcaxy distribution in the range 0.55 $<$ \pt\xspace$<$ 0.60\gevc together with the Monte-Carlo templates for primary protons (green dotted line), secondary protons from weak decays (red dotted line) and secondary protons produced in interactions with the detector material (blue dashed line) which are fitted to the data. The light blue line represents the combined fit, while the black dots are the data.}
\label{fig:DCA}
\end{figure}

The contamination due to weak decays of light flavour hadrons (mainly K$^0_s$ affecting $\pi$ spectra and $\Lambda$ and $\Sigma^{+}$ affecting p spectra) and interactions with the material has to be computed and subtracted from the raw spectra. 
Since strangeness production is underestimated in the event generators and the interactions of low \pt \xspace particles with the material are not properly modelled in the transport codes, the secondary-particle contribution is evaluated with a data-driven approach.
This approach exploits the high resolution determination of the track impact parameter in the transverse plane, \dcaxy, and the fact that secondary particles from strange hadron decays and interactions with the detector material, originate from secondary vertices significantly displaced from the interaction point and, therefore, their tracks have, on average, larger absolute values of \dcaxy with respect to primary particles. 
Hence, for each of the PID techniques described in the previous sections, the contribution of secondary particles to the measured raw yield of a given hadron species in a given \pt\xspace interval is extracted by fitting the measured distributions of \dcaxy of the tracks identified as particles of the considered hadron species.
The \dcaxy distributions are modelled with three contributions, called templates.
Their shapes are extracted for each \pt\xspace interval and particle species from simulations, as described in~\cite{centralityPbPb}, and represent the \dcaxy distributions of primary particles, secondary particles from weak decays of strange hadrons and secondary particles produced in interactions with the detector material, respectively. 
An example for protons in the interval 0.55 $<$ \pt\xspace$<$ 0.60\gevc is shown in Fig.~\ref{fig:DCA}.
\begin{figure}
\centering
\includegraphics[width=16.0cm]{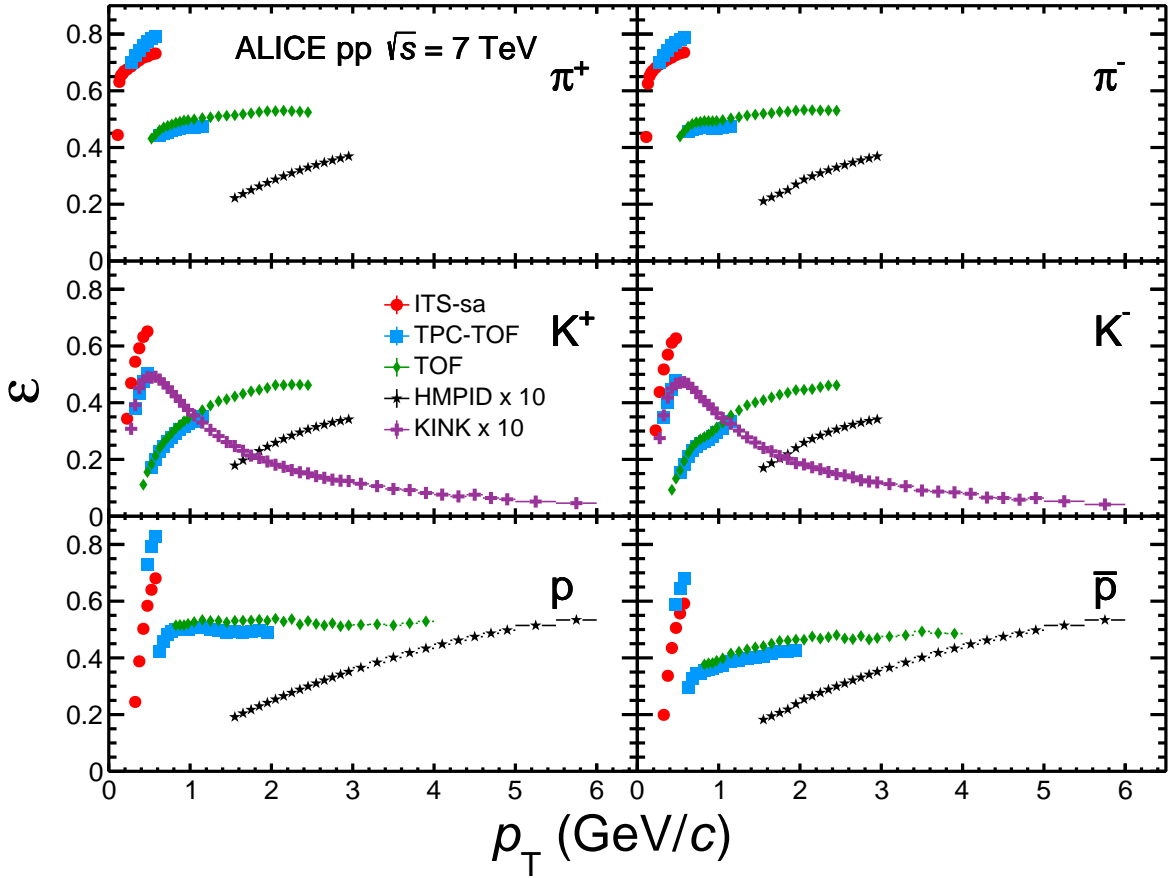}
\caption{Correction factors ($\varepsilon$(\pt) in Eq. \ref{eq:corr}) for $\pi^{+}$, K$^{+}$ and p (left panel) and their antiparticles (right panel) accounting for PID efficiency, detector acceptance, reconstruction and selection efficiencies for ITS-sa (red circles), TPC-TOF (light blue squares), TOF (green diamonds), HMPID (black stars) and kink (purple crosses) analyses.}
\label{fig:totcorr}
\end{figure} 
The correction for secondary particle contamination is relevant for $\pi^\pm$ (from 10\% at low \pt\xspace to less than 2\% at high \pt), p and \pbar (from 35\% at low \pt\xspace to 2\% at high \pt).
Due to the different track and PID selections the contribution of secondaries is different for each analysis.  

In the case of kaons, the contamination from secondary particles is negligible, except for the TPC-TOF analysis where a contamination originating from secondary $\rm{e}^{\pm}$ produced by photon conversions in the detector material is present. This contamination is significant only in the momentum range 0.4~$<$~$p$~$<$~0.6~\gevc, where the \dedx of kaons and electrons in the TPC gas are similar, not allowing for their separation, as shown in Fig. \ref{fig:TPC:dEdx}. Therefore, in the case of kaons, the fit to the DCA$_{xy}$ distributions is used only in the TPC-TOF analysis for \pt \xspace$ < $ 0.5\gevc to subtract the contamination due to secondary $\rm{e}^{\pm}$. This contamination is about 16\% for \pt\xspace= 0.5\gevc.

The resulting spectra are corrected for the detector acceptance and for the reconstruction and selection efficiencies. This correction is specific to each analysis and accounts for the acceptance of the detector used in the PID procedure, the trigger selection and the vertex and track reconstruction efficiencies.
They are evaluated by performing the same analyses on simulated events generated with PYTHIA 6.4 (Perugia0 tune) \cite{PYTHIA-P2011}. The particles are propagated through the detector using the GEANT3 transport code~\cite{GEANT3}, where the detector geometry and response as well as the data taking conditions are reproduced in detail. 

In Fig.~\ref{fig:totcorr} the efficiency $\varepsilon$(\pt), specific to each analysis, accounting for PID efficiency, acceptance, reconstruction and selection  efficiencies are shown.
The lower value of $\varepsilon$ for HMPID and kink analyses 
is due to the limited geometrical acceptance of the HMPID detector and to the limited TPC fiducial volume used for the kink vertex reconstruction.
The drop in the correction for the TPC-TOF analysis at \pt\xspace= 0.6~\gevc for pions and protons and \pt\xspace= 0.5\gevc for kaons is due to the efficiency of track propagation to the TOF. 
The ITS-sa analysis has a larger kaon efficiency than TPC-TOF analysis at low \pt\xspace because the ITS-sa tracking allows the reconstruction of kaons that decay before reaching the TPC. The corrections for particles (left panel of Fig.~\ref{fig:totcorr}) and antiparticles (right panel) are compatible within the uncertanties.

Since GEANT3 does not describe well the interaction of low-momentum \pbar and K$^{-}$ with the material, corrections to the efficiencies, estimated with a dedicated FLUKA simulation~\cite{centralityPbPb,FLUKA}, are applied. The correction factor $f$(\pt) is 0.71~$<$~$f$(\pt)~$<$~1 for \pbar and 0.95~$<$~$f$(\pt)~$<$~1 for K$^{-}$.

The corrected spectra are, finally, normalized to the number of inelastic proton-proton collisions that is obtained from the number of analyzed minimum-bias events via the scaling factor 0.852 as described in~\cite{ALICE_INEL}.

\subsection{Systematic uncertainties}
\begin{table}[t]
\centering

\begin{tabular}{lcccccc}
\hline
\hline
 \textbf{Source of uncertainty common to all the analyses}    &  \multicolumn{2}{c}{$\pi^\pm$}  &   \multicolumn{2}{c}{K$^\pm$}  &  \multicolumn{2}{c}{p and \pbar}   \\ 
\hline
\hline
 Correction for secondaries & \multicolumn{2}{c}{$<1\%$} & \multicolumn{2}{c}{} & \multicolumn{2}{c}{$5\% - 1.5\%$~(p)} \\
 &  &  &  \multicolumn{2}{c}{ }  &  \multicolumn{2}{c}{1.5\%~(\pbar)} \\
Material budget & \multicolumn{2}{c}{$5\% - \mathrm{Negl.}$} & \multicolumn{2}{c}{$3\% - \mathrm{Negl.}$} & \multicolumn{2}{c}{$3\% - \mathrm{Negl.}$} \\
Cross sections for interactions in the material  & \multicolumn{2}{c}{$2\%- 1\%$} & \multicolumn{2}{c}{$4\% - 1\%$}  & \multicolumn{2}{c}{$4\% - \mathrm{Negl.}$~(p)}\\
 &  &  &  &  & \multicolumn{2}{c}{$6\% - 1\%$~(\pbar)}\\
ITS-TPC matching (excluded in ITS-sa analysis)  & \multicolumn{2}{c}{3\%-} & \multicolumn{2}{c}{3\%} & \multicolumn{2}{c}{3\%} \\ 
\hline
\hline
\textbf{Source of uncertainty specific to an analysis}    &  \multicolumn{2}{c}{$\pi^\pm$}  &   \multicolumn{2}{c}{K$^\pm$}  &  \multicolumn{2}{c}{p and \pbar}   \\ 
\hline
\hline
ITS-sa PID  & \multicolumn{2}{c}{2\%} & \multicolumn{2}{c}{4\%} & \multicolumn{2}{c}{4.5\%} \\ 
Tracking efficiency (ITS-sa tracks) & \multicolumn{2}{c}{3\%} & \multicolumn{2}{c}{3\%} & \multicolumn{2}{c}{3\%} \\
$\it{E} \rm{\times} \it{B}$ effect & \multicolumn{2}{c}{3\%} & \multicolumn{2}{c}{3\%} & \multicolumn{2}{c}{3\%} \\
\hline
TPC-TOF PID  & \multicolumn{2}{c}{$<1\%$} & \multicolumn{2}{c}{$1\% -  5\%$}  & \multicolumn{2}{c}{$<1\%$} \\ 
Tracking efficiency (global tracks)  & \multicolumn{2}{c}{2\%} & \multicolumn{2}{c}{2\%} & \multicolumn{2}{c}{2\%} \\ 
Matching efficiency & \multicolumn{2}{c}{3\%} & \multicolumn{2}{c}{6\%} & \multicolumn{2}{c}{4\%} \\ 
(\pt\xspace$>$ 0.5\gevc for K and 0.6\gevc for $\pi$, p) & \multicolumn{6}{c}{} \\
\hline
TOF PID  & \multicolumn{2}{c}{$0.5\% -  3\%$} & \multicolumn{2}{c}{$1\% - 11\%$} & \multicolumn{2}{c}{$1\% -  14\%$} \\ 
Tracking efficiency (global tracks)  & \multicolumn{2}{c}{2\%} & \multicolumn{2}{c}{2\%} & \multicolumn{2}{c}{2\%} \\ 
Matching efficiency  & \multicolumn{2}{c}{3\%} & \multicolumn{2}{c}{6\%} & \multicolumn{2}{c}{4\%} \\ 
\hline
HMPID PID  & \multicolumn{2}{c}{4\%} & \multicolumn{2}{c}{5\%}  & \multicolumn{2}{c}{$5\% - 9\%$} \\
Tracking efficiency (global tracks)  & \multicolumn{2}{c}{5\%} & \multicolumn{2}{c}{5\%} & \multicolumn{2}{c}{7\%} \\ 
$\rm{d}_{\rm{MIP-trk}}$ cut & \multicolumn{2}{c}{$2\% - 6\%$} & \multicolumn{2}{c}{$2\% - 6\%$} & \multicolumn{2}{c}{$2\% - 6\%$} \\ 
\hline
Kink PID   & \multicolumn{2}{c}{} & \multicolumn{2}{c}{3\%} & \multicolumn{2}{c}{} \\
Tracking efficiency (global tracks)  & \multicolumn{2}{c}{} & \multicolumn{2}{c}{2\%} & \multicolumn{2}{c}{} \\ 
Kink reconstruction efficiency   & \multicolumn{2}{c}{} & \multicolumn{2}{c}{3\%} & \multicolumn{2}{c}{} \\
Kink contamination   & \multicolumn{2}{c}{} & \multicolumn{2}{c}{$3.6\% - 2\%$} & \multicolumn{2}{c}{} \\

\hline
\hline
\end{tabular}
\caption{Sources of systematic uncertainties on the corrected spectra $\frac{{\rm d}^{2} N}{{\rm d} p_{\rm{T}} {\rm d} y}$. In case of \pt-dependent systematic uncertainty, the values in the lowest and highest \pt\xspace intervals are reported.}
\label{tab:systematics}
\end{table}

The main sources of systematic uncertainties, for each analysis, are summarized in Table~\ref{tab:systematics}. They are related to the PID procedure, the subtraction of the contribution from secondary particles, imperfect description of the material budget in the Monte-Carlo simulation, particle interactions in the detector material, tracking efficiency and variables used for the track selection.

The systematic uncertainties introduced by the PID procedure are estimated differently depending on the specific analysis.
In the ITS-sa analysis different techniques are used for the identification: a 3$\sigma$ compatibility cut and an unfolding method as described in section \ref{sec:itssa}. 
In the TPC-TOF analysis the 3$\sigma$ selection is varied to 2$\sigma$ and 4$\sigma$.
Furthermore, the systematic uncertainty on the estimated contamination from misidentified hadrons, which is due to the different relative abundances of pions, kaons and protons in data and simulation, has been estimated to be below 1\% for pions and protons and below 4\% for kaons.
In TOF and HMPID analyses the parameters of the fit function used to extract the raw yields are varied (one at a time) by $\pm$10\%. 

The systematic uncertainty due to the subtraction of secondary particles is estimated by changing the fit range of the \dcaxy distribution. The shape of the \dcaxy template for p and \pbar from weak decays is also varied by modifying the relative contribution of the different mother particles.
The main sources of p and \pbar from weak decays are $\Lambda$ and $\Sigma^{+}$ (and their antiparticles), which have significantly different mean proper decay lengths ($c\tau$ = 7.89 cm and 2.404 cm respectively~\cite{PDG}). Therefore, the DCA template of protons from weak decays depends on the  $\Lambda$ to $\Sigma^{+}$ ratio in the event generator used in the simulation.   

To evaluate the systematic effect due to the uncertainty in the material budget (about $\pm$7\%~\cite{mat_budget}),
the efficiency corrections are computed by using Monte-Carlo simulations with the material budget modified by this percentage.
The systematic uncertainties in modelling the particle interactions with the detector material are evaluated using different transport codes, as described in~\cite{centralityPbPb}.

For all the analyses, the systematic uncertainties related to tracking procedure are estimated by varying the track selection criteria (e.g. number of crossed readout rows in TPC, number of clusters in ITS,  \dcaz, \dcaxy) reported in Section \ref{subsec:evtsel}.
For global tracks an additional uncertainty, related to the ITS-TPC matching, is also included. 
It is estimated by comparing the matching efficiency in data and Monte-Carlo simulations. 

Further systematic uncertainty sources, specific to each analysis, are also evaluated.
In the case of the ITS-sa analysis, the Lorentz force causes shifts of the cluster position in the ITS, pushing the charge in opposite directions depending on the polarity of the magnetic field of the experiment ($E$$\rm{\times}$$B$ effect). 
This effect is not fully reproduced in the simulation. It is estimated by analysing data samples collected with different magnetic field polarities, which resulted in an uncertainty of 3\%. 
In the case of TPC-TOF and TOF analyses, the influence of the material budget on the matching of global tracks with hits in the TOF detector is computed by comparing the matching efficiency for tracks traversing a different amount of material, in particular sectors with and without Transition Radiation Detector (TRD) modules installed. 
In the HMPID analysis, the $d_{\rm{MIP-trk}}$ cut selection is varied to check its systematic effect on the matching of global tracks with HMPID signals.

In the kink analysis, the total systematic uncertainty is calculated as the quadratic sum of the contributions listed in Table~\ref{tab:systematics}. The  kaon misidentification correction (1~-~purity) described in Section~\ref{kink_analysis}, which is on average 2.1\%, depends on the relative particle abundances in the Monte-Carlo  and a  \pt-dependent uncertainty of about 2\% on the purity is estimated.
The kink identification uncertainty (3\%, almost flat in the considered \pt\xspace region) is also estimated with Monte-Carlo simulations by comparing the results by varying slightly some parameters of the analysis: the fiducial volume of the TPC is increased from the nominal 130 $ < R < $ 200 cm to 120 $ < R < $ 210 cm, the $q_\mathrm{T}$ threshold is reduced from the nominal 120 \mevc to 40 \mevc , and the requirement on the number of TPC clusters of the mother track is increased from the nominal 20 to 50 clusters.\\
The systematic uncertainty on the efficiency for findable kink vertices was estimated to be 3\% independently of \pt\xspace by comparing, in real data and Monte Carlo simulations, the number of raw reconstructed kinks per kink radius unit in two different fiducial volumes inside the TPC, namely 130-200 cm and 140-190 cm.

Finally, a systematic uncertainty common to each analysis is related to the normalization to inelastic collisions. The normalization factor was evaluated in~\cite{ALICE_INEL} and it is 0.852$^{+0.062}_{-0.030}$.  

All described uncertainties are strongly correlated among the \pt\xspace bins.
Most of the uncertainties (e.g. tracking efficiency, ITS-TPC matching, TOF matching, material budget or PID) are also correlated between the different particle species.  

\begin{figure}[t]
\centering
\includegraphics[trim=0cm 0.cm 0cm 0cm, clip=true, width=8.5cm]{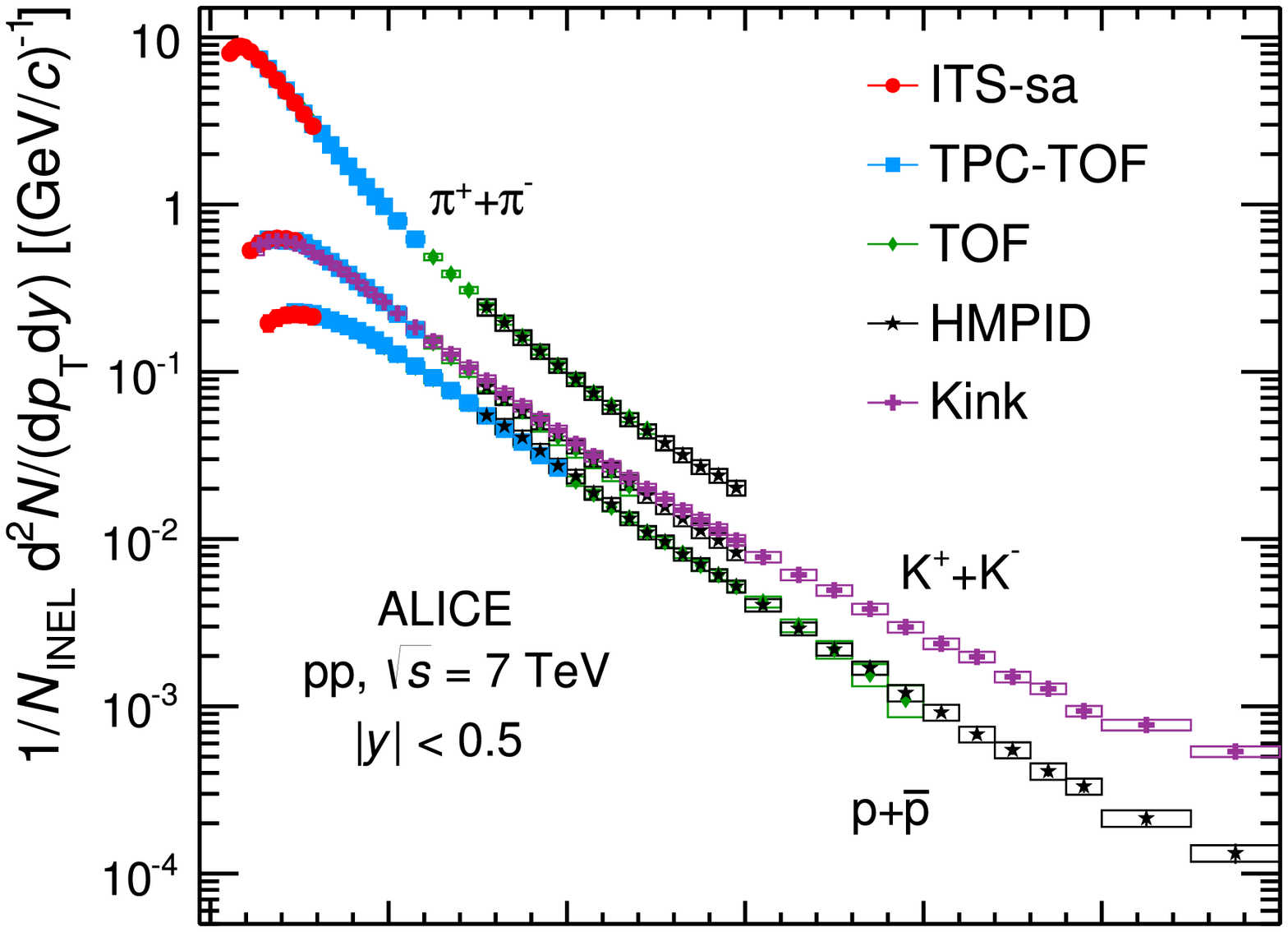} \hspace{1.5cm}
\includegraphics[trim=0cm 0cm 0cm 0.7cm, clip=true, width=8.5cm]{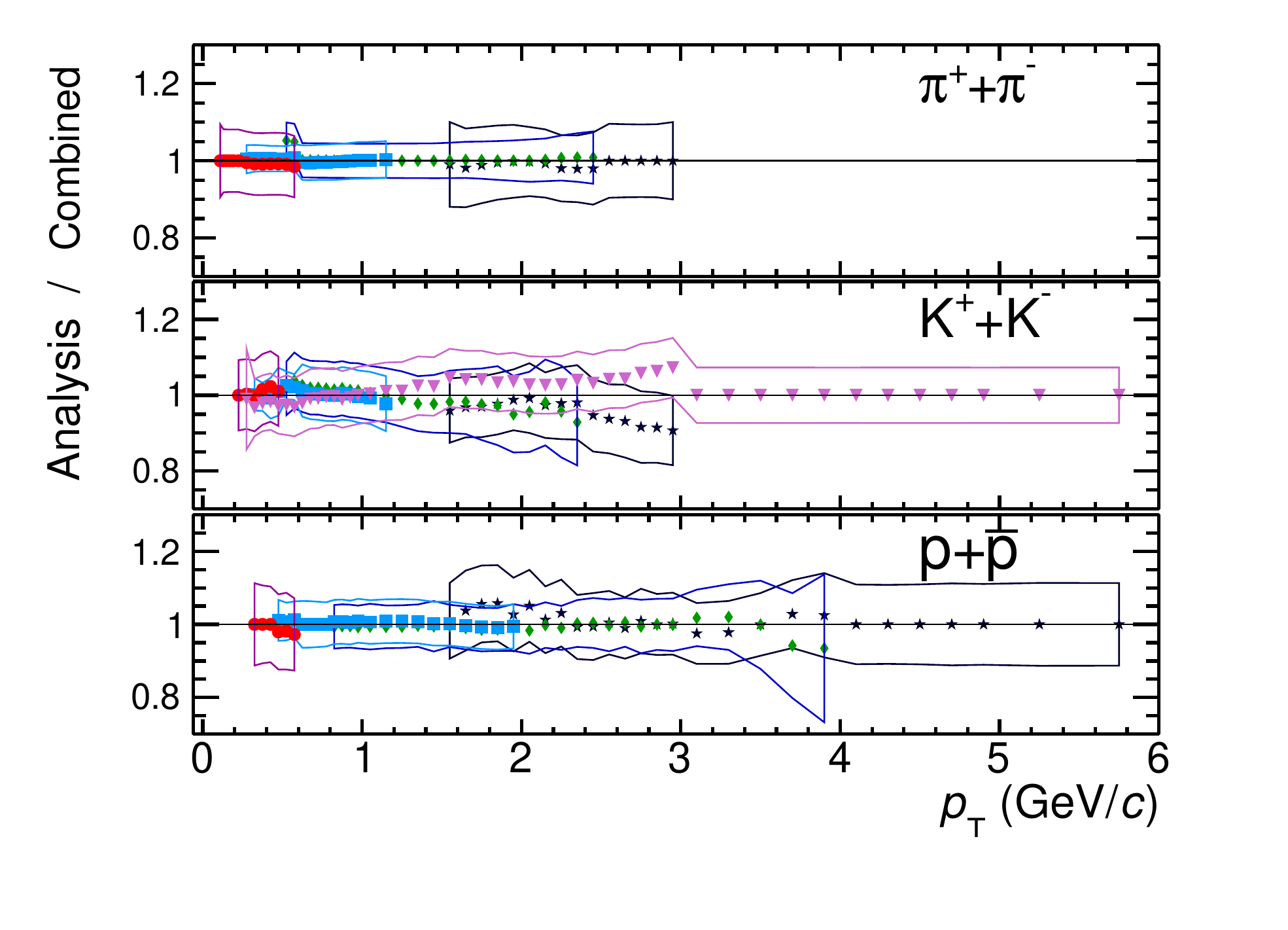}
\caption{Top panel: \pt\xspace spectra of $\pi$, K and p, sum of particles and antiparticles, measured with ALICE at mid-rapidity ($|y| <$ 0.5) in pp collisions at $\sqrt{s}=$ 7 TeV by using different PID techniques. The spectra are normalized to the number of inelastic collisions. Statistical (vertical error bars) and systematic (open boxes) uncertainties are reported. The horizontal width of the boxes represents the \pt-bin width.
 The markers are placed at the bin centre.
Bottom panels: ratio between the spectra obtained from each analysis and the combined one. The error bands represent the total systematic uncertainties for each analysis. The uncertainty due to the normalization to inelastic collisions ($ ^{+7}_{-4} \%$), common to the five PID analyses, is not included.}
\label{fig:COMB:comp}
\end{figure} 

\section{Results}
\begin{figure}[t]
\centering
\includegraphics[width=8.5cm]{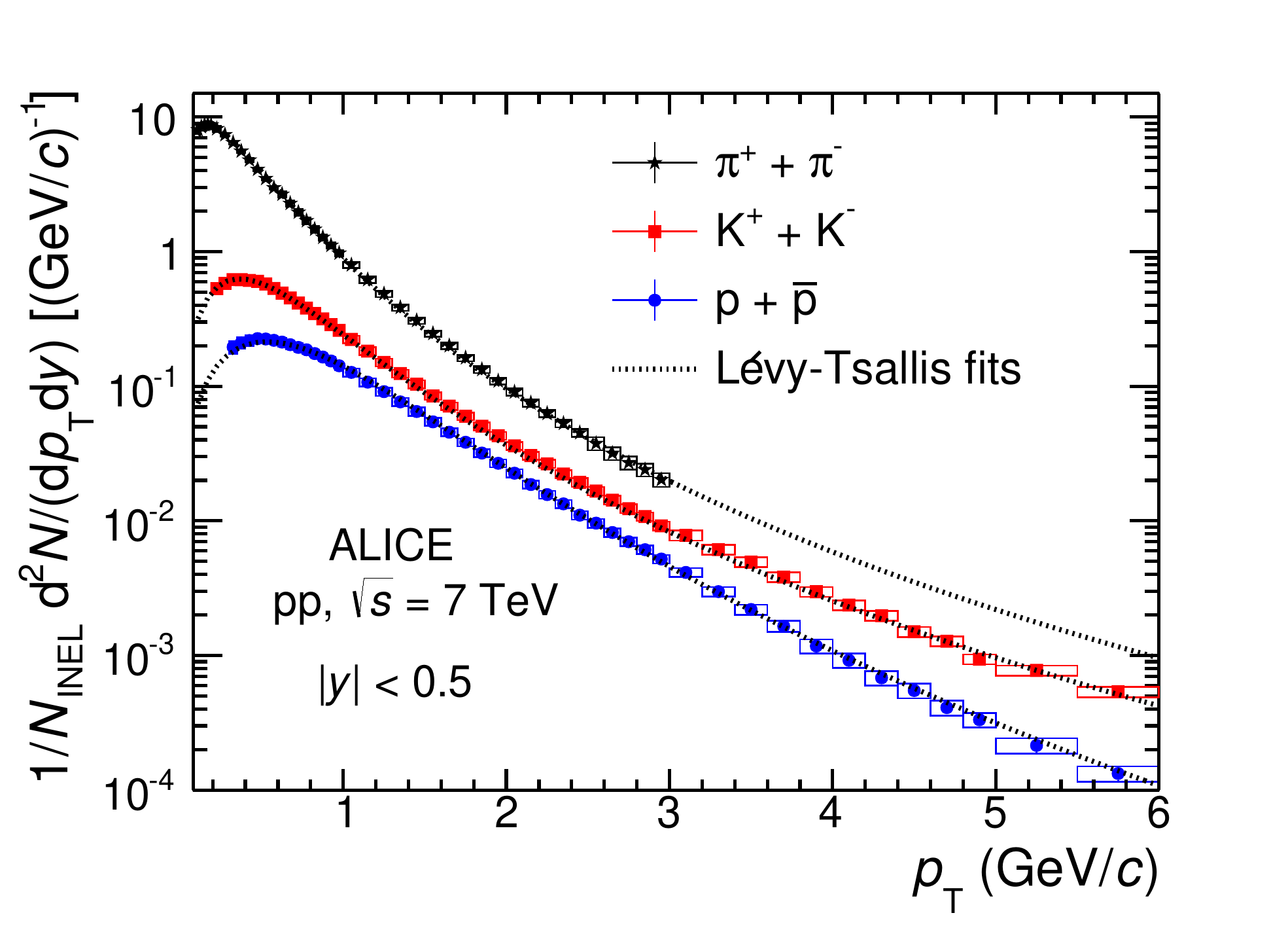}
\caption{Combined \pt\xspace spectra of $\pi$, K and p, sum of particles and antiparticles, measured with ALICE at mid-rapidity ($|y| <$ 0.5) in pp collisions at $\sqrt{s}=$ 7 TeV normalized to the number of inelastic collisions. Statistical (vertical error bars) and systematic (open boxes) uncertainties are reported. The uncertainty due to the normalization to inelastic collisions ($ ^{+7}_{-4} \%$) is not shown. The spectra are fitted with L\'evy-Tsallis functions.}
\label{fig:COMB:fit}
\end{figure}

The mid-rapidity ($|y| <$ 0.5) transverse momentum spectra of $\pi^{+}$+$\pi^{-}$, K$^{+}$+K$^{-}$ and p+\pbar obtained with the five analysis techniques discussed in Section~\ref{sec:analysis}, normalized to the number of inelastic collisions $N_{\rm{INEL}}$, are reported in the top panel of Fig.~\ref{fig:COMB:comp}. 
For a given hadron species, the spectra of particles and antiparticles  are found to be compatible within uncertainties.
Therefore, all the spectra shown in this section are reported for summed charges.
Since in their overlap \pt\xspace regions the spectra from the different PID techniques are consistent within uncertainties, they are averaged in a sequential procedure.
The first step consists in averaging the two analyses whose results are the most closely correlated (namely TPC-TOF and TOF).
Successively, the other analyses are added one-by-one to the running average according to their degree of correlation with the previous ones.
At each step of this sequential procedure, a weighted average of two spectra is computed by using as weights the inverse of the squares of the uncorrelated systematic uncertainties. The uncorrelated and correlated uncertainties are propagated separately through the weighted average formula.
In Fig.~\ref{fig:COMB:fit} the $\pi$, K and p spectra, resulting from the combination of the five analyses, are reported. 
The bottom panels of Fig.~\ref{fig:COMB:comp} show the ratios between the spectra from each analysis and the combined one: the former are considered with their total systematic uncertainties while the latter without uncertainty.  The uncertainty due  to the normalization to inelastic collisions ($ ^{+7}_{-4} \%$), common to the five PID analyses, is not included. The agreement between each analysis and the combined one is satisfactory, being within the total systematic uncertainties. 

To extrapolate to zero and infinite momentum, the combined spectra reported in  Fig.~\ref{fig:COMB:fit} are fitted with the L\'evy-Tsallis function~\cite{tsallis1,tsallis2}
\begin{equation}
\frac{{\rm d}^{2} N}{{\rm d} p_{\rm{T}} {\rm d} y} = p_{\rm{T}} \frac{{\rm d} N}{{\rm d} y} K \left(  1 + \frac{m_{\rm{T}} - m_{0}}{n C} \right)^{-n},
\end{equation}
where
\begin{equation}
K = \frac{(n - 1) (n - 2)}{n C (n C + m_{0} (n - 2))}  \, \,\,\,\,\xspace ,
\end{equation}
$m_{\rm{T}} = \sqrt{p_{\rm{T}}^{2} + m_{0}^{2}}$, $m_{0}$ is the particle rest mass and $C$, $n$ and the yield \dndy are the free parameters.
The L\'evy-Tsallis function describes rather well the spectra. The $\chi^{2}$ per number of degrees of freedom (ndf) of the fit are lower than unity (see Table \ref{tab:comb:yieldpt}) due to residual correlations in the point-to-point systematic uncertainties.
In Table \ref{tab:comb:yieldpt} the values of the \pt\xspace-integrated yield \dndy and of the mean transverse momentum \ptmean \xspace are reported for each particle species. They are obtained using the measured data in the \pt\xspace range where they are available and the L\'evy-Tsallis function fitted to the data elsewhere, to extrapolate to zero and infinite momentum.
The lowest \pt \xspace  experimentally accessible and the fraction of yield contained in the extrapolated region are also reported in the table.
The extrapolation to infinite momentum gives a negligible contribution to the values of both \dndy and \ptmean. 
The \dndy and \ptmean\xspace uncertainties reported in Table \ref{tab:comb:yieldpt} are the combination of the statistical and the systematic ones. The statistical uncertainties are negligible, while the systematic uncertainties are the sum of two independent contributions.  
The first contribution is due to the systematic uncertainties on the measured \pt-differential yields and it was estimated 
by repeating the L\'evy-Tsallis fits moving the measured points within their systematic uncertainties.
The second contribution is due to the extrapolation to zero momentum and it is  estimated using different fitting functions (namely modified Hagedorn~\cite{Hagedorn1984} and UA1 parametrization~\cite{UA1}). Results for positively and negatively charged particles, separately, are also reported. It should be noticed that the yields of particles and antiparticles are compatible within uncertainties.

\begin{table}[t]
\centering
\begin{tabular}{cccccc}
\hline
\hline
\textbf{Particle} & \textbf{d}$\mathbf{\textit{N}}/$\textbf{/d}$\mathbf{\textit{y}}$&\head{2.0cm}{\textbf{$\mathbf{\langle \textit{p}_{\mathbf{T}} \rangle}$ (GeV/$\mathbf{c}$)}} &$\mathbf{\chi^{2}/ndf}$& \head{1.0cm}{\textbf{L. $\mathbf{\textit{p}_{\mathbf{T}}}$ (GeV/$\mathbf{c}$)}} &\head{1.0cm}{\textbf{Extr. \%} }\\
\hline 
\hline
$\pi^{+} + \pi^{-}$ & $4.49 \pm 0.20$ &  $0.466
\pm 0.010 $ & 19.1/38 & 0.10  & 9 \\
K$^{+}$ + K$^{-}$& $ 0.572 \pm 0.032 $ &  $0.773
\pm 0.016 $ & 5.0/45 & 0.20  & 10 \\
p + \pbar & $0.247 \pm 0.018 $ &  $0.900  \pm 0.029
$ & 10.8/43 & 0.30 & 12 \\
\hline
$\pi^{+}$ & $2.26  \pm 0.10$ &  $0.464 \pm 0.010$ &
24.0/38 & 0.10  & 9 \\
$\pi^{-}$ & $2.23  \pm 0.10$ &  $0.469  \pm 0.010$ &
15.0/38 & 0.10  & 9 \\
K$^{+}$ & $0.286  \pm 0.016$ &  $0.777  \pm 0.016$ &
7.4/45 & 0.20 & 9 \\
K$^{-}$ & $0.286  \pm 0.016$ &  $0.770  \pm 0.016$ &
10.0/45 & 0.20 & 10 \\
p & $0.124 \pm 0.009$ &  $0.900 \pm 0.027$ & 9.5/43
& 0.30  & 12 \\
$\overline{\rm p}$ & $0.123  \pm 0.010$ &  $0.900 \pm
0.032$ & 12.3/43 & 0.30  & 12 \\
\hline
\end{tabular}
\caption{\dndy and \ptmean~extracted from L\'evy-Tsallis fits to the measured $\pi$, K, p spectra in inelastic pp collisions at \sqrts = 7 TeV for $|y|<0.5$ with combined statistical and systematic uncertainties (statistical uncertainties are negligible) together with the \pt \xspace of the lowest experimentally accessible point (L. $\textit{p}_{\mathbf{T}}$) and the extrapolated fraction. The systematic uncertainty on \dndy due to normalization to inelastic collisions ($ ^{+7}_{-4} \%$) is not included.}
\label{tab:comb:yieldpt}
\end{table}

\begin{figure}
\centering
\includegraphics[width=8.5cm]{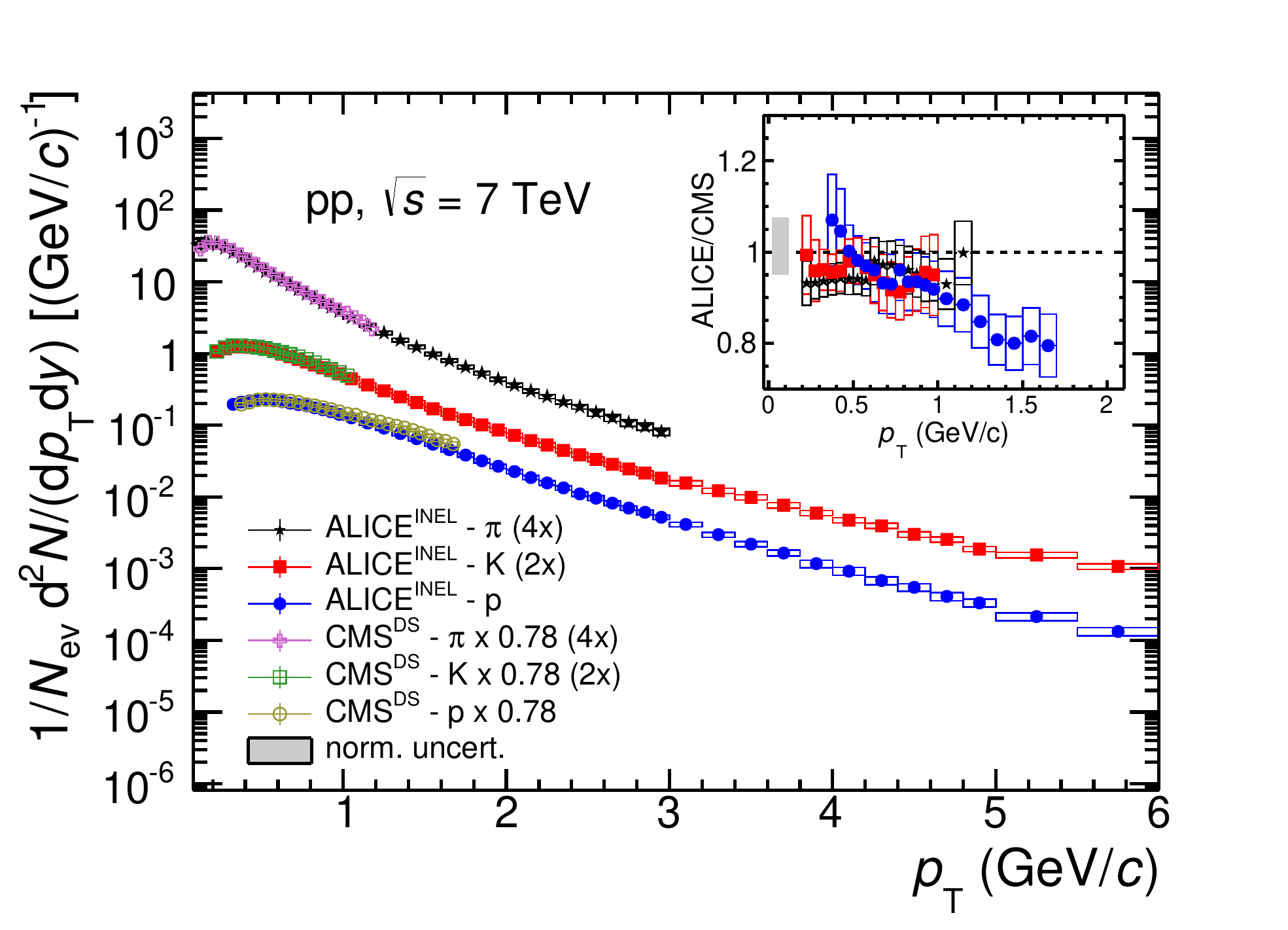}
\caption{Comparison of \pt\xspace spectra of $\pi$, K and p (sum of particles and antiparticles) measured by the ALICE ($|y|<0.5$) and CMS Collaborations ($|y|<1$) in pp collisions at \sqrts = 7 TeV. The CMS data points are scaled by the empirical factor 0.78, as described in~\cite{CMS_spectra}. Inset plot: Ratios between ALICE and CMS data in the common \pt\xspace range. The combined ALICE and CMS statistical (vertical error bars) and systematic (open boxes) uncertainties are reported. The combined ALICE ($ ^{+7}_{-4} \%$) and CMS ($\pm 3\%$) normalization uncertainty is shown as a grey box around 1 and not included in the point-to-point uncertainties.
}
\label{fig:ALICE:CMS}
\end{figure}

In Fig.~\ref{fig:ALICE:CMS} the \pt\xspace spectra of identified charged hadrons, sum of particles and antiparticles, measured with ALICE at $\sqrt{s}=$ 7 TeV are compared to the results obtained by the CMS Collaboration at the same centre-of-mass energy~\cite{CMS_spectra}. Even though the measurements are performed in different rapidity intervals ($|y|<0.5$ for ALICE, $|y|<1$ for CMS), they can be compared since the \pt\xspace spectra are essentially independent of rapidity for $|y|<1$. 
A similar comparison at \sqrts = 0.9 TeV is reported in~\cite{CMS_spectra}. At both energies, the ALICE spectra are normalized to the number of  inelastic collisions, while the CMS results are normalized to the double-sided selection (at least one particle with $E > 3$ GeV in both $-5 < \eta < -3$ and $3 < \eta < 5$). An empirical scaling factor of 0.78, computed by the CMS Collaboration in ~\cite{CMS_spectra} for the spectra measured in  pp collisions at  \sqrts = 0.9 TeV, is therefore applied to the CMS data points at \sqrts = 7 TeV, to take into account the different event selections (details are given in~\cite{CMS_spectra}). With this scaling, the  pion and kaon spectra measured with ALICE and CMS are found to agree within uncertainties.
The proton spectra have different slopes:  for  \pt\xspace$<$ 1\gevc the ALICE and CMS results agree within uncertainties, while at higher  \pt\xspace a discrepancy of up to 20\% is observed.
 
\begin{figure}
\centering
\includegraphics[width=8.5cm]{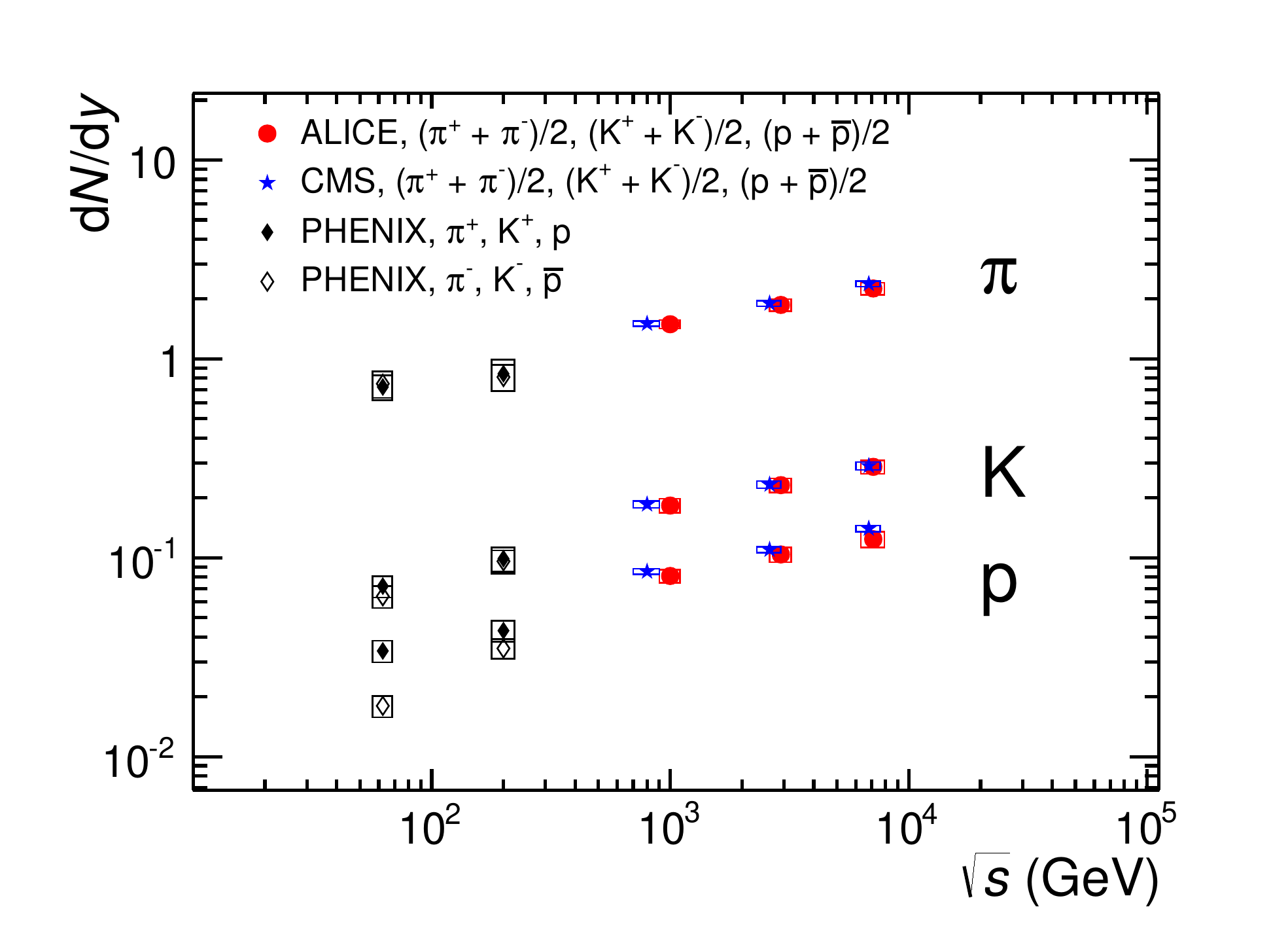}
\caption{\pt-integrated yields \dndy of $\pi$, K and p as a function of the centre-of-mass energy in pp collisions. PHENIX results are for separate charges while CMS and ALICE results are the average of the \dndy of particles and antiparticles. ALICE and CMS points are slightly shifted along the x-axis for a better visualization. Errors (open boxes) are the combination of statistical (negligible), systematic and normalization uncertainties.}
\label{fig:dNdY}
\end{figure}

\begin{figure}
\centering
\includegraphics[width=8.5cm]{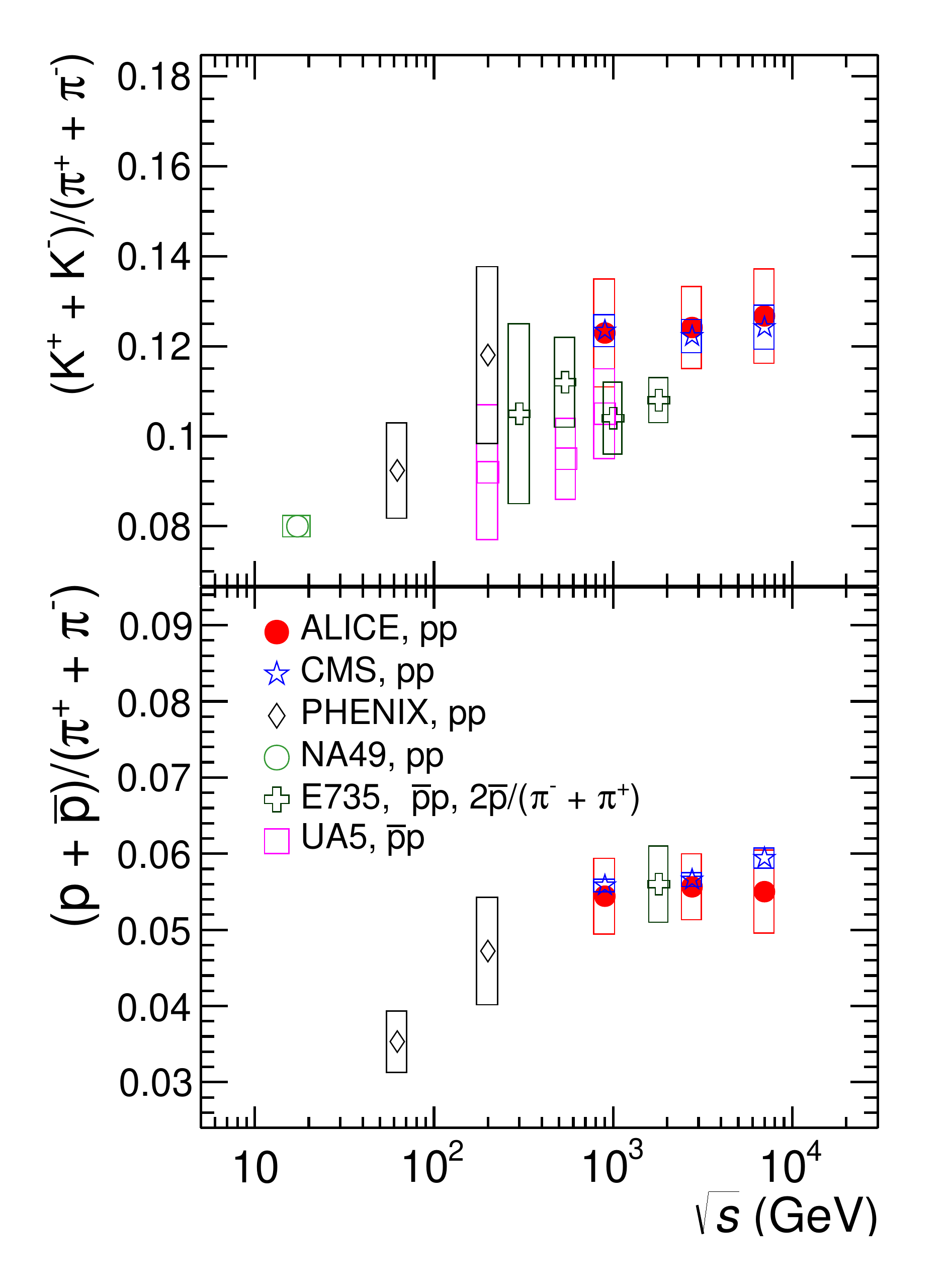}
\caption{(K$^{+}$+K$^{-}$)/($\pi^{+}$+$\pi^{-}$) (top) and (p+\pbar)/($\pi^{+}$+$\pi^{-}$) (bottom) ratios in pp and p$\overline{\mathrm{p}}$ collisions as a function of the collision energy \sqrts. 
Errors (open boxes) are the combination of statistical (negligible) and systematic uncertainties.
}
\label{fig:RATIO:yield}
\end{figure}

\begin{figure}[h]
\centering
\includegraphics[width=8.5cm]{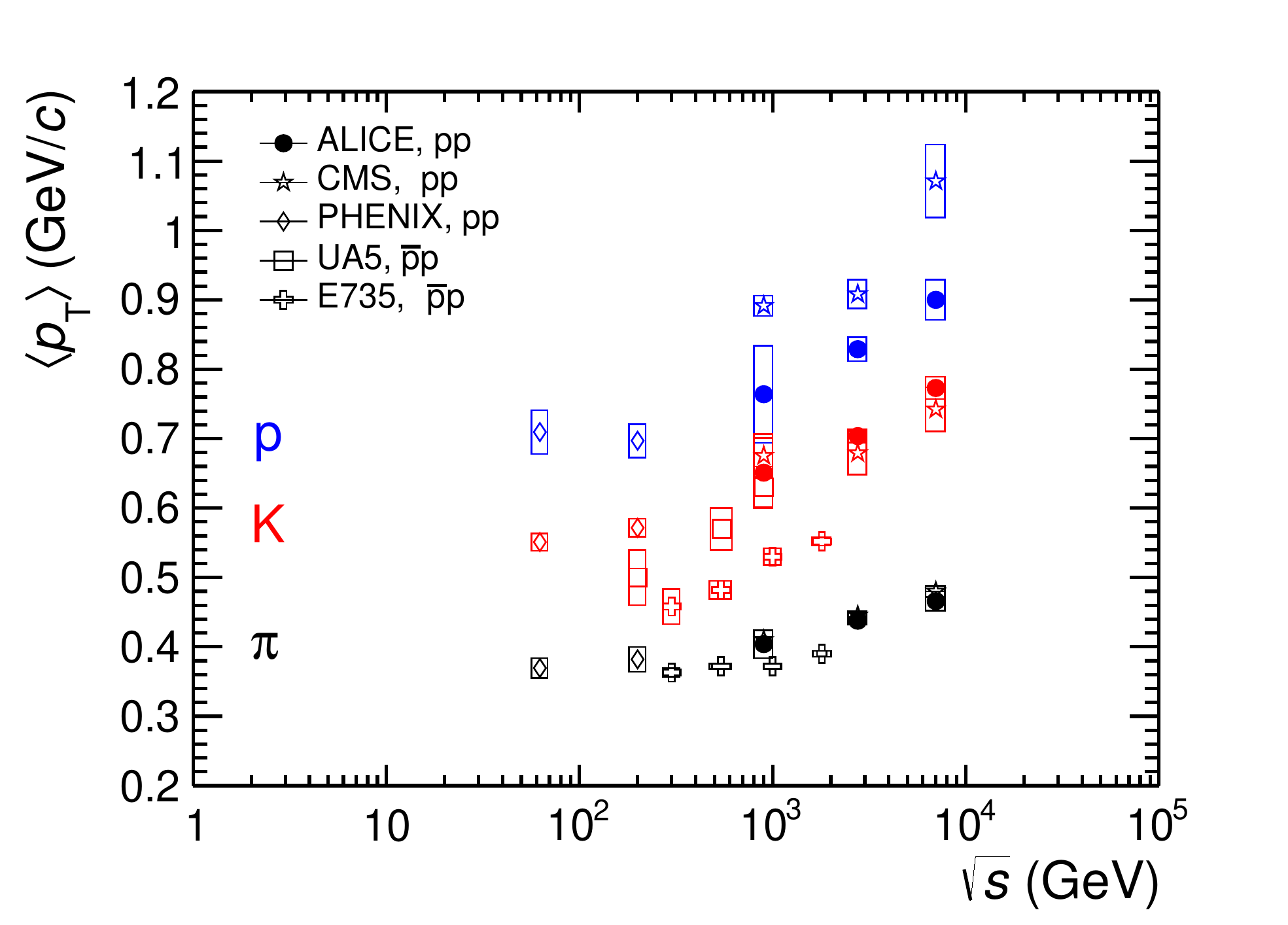}
\caption{(Colour online) \ptmean~as a function of the centre-of-mass energy. 
Errors (open boxes) are the combination of statistical (negligible) and systematic uncertainties. Normalization uncertainties are not included.}
\label{fig:MEANPT:mass}
\end{figure}

In Fig.~\ref{fig:dNdY} the $\pi$, K and p integrated yields, \dndy, are compared with similar measurements in the central rapidity region at various collision energies. In particular, results  from ALICE at \sqrts = 900 GeV~\cite{900GeVspectra} and \sqrts = 2.76 TeV~\cite{276ALICESpectra}, PHENIX at \sqrts = 62.4 GeV and \sqrts = 200 GeV~\cite{PHENIX_2010} and CMS, scaled by the empirical factor 0.78, at \sqrts = 900 GeV, \sqrts = 2.76 TeV and  \sqrts = 7 TeV~\cite{CMS_spectra} are shown.  The \dndy  values from  PHENIX  are reported for particles and antiparticles separately, while the results at LHC energies are the average between  positively and negatively  charged particles, since particle and antiparticle spectra are compatible  at these energies. We notice that the CMS Collaboration does not include, in the systematic uncertainties associated to \dndy and \ptmean, the contribution due to the extrapolation to \pt~=~0. For this reason,  in Fig.~\ref{fig:RATIO:yield} and Fig.~\ref{fig:MEANPT:mass}, the ALICE uncertainties are larger than the  CMS ones.  Similar results from the STAR Collaboration ~\cite{STAR_2009} are not included, here and in the following plots, since they are provided for non-single diffractive events and include contributions of feed-down from weak decays.

The (K$^{+}$+K$^{-}$)/($\pi^{+}$+$\pi^{-})$ and (p+\pbar)/($\pi^{+}$+$\pi^{-}$) ratios, as a function of the centre-of-mass energy, are shown in the top and bottom panels of Fig.~\ref{fig:RATIO:yield}, respectively.
Results at mid-rapidity from ALICE at \sqrts = 0.9, 2.76~\cite{900GeVspectra, 276ALICESpectra} and 7 TeV, CMS at \sqrts = 0.9, 2.76 and 7 TeV~\cite{CMS_spectra}, PHENIX at \sqrts = 62.4 and 200 GeV~\cite{PHENIX_2010} and NA49 at \sqrts = 17.3 GeV~\cite{NA49r1,NA49r2,NA49r3} are displayed.
The ratio (p+\pbar)/($\pi^{+}$+$\pi^{-}$) from NA49, calculated from the measured particle yields, is not reported because the uncertainty can not be computed from the results published in~\cite{NA49r1,NA49r2,NA49r3}. Results in proton-antiproton collisions from E735 at \sqrts = 0.3, 0.54, 1 and 1.8 TeV~\cite{e735,E735prompt} and UA5 at \sqrts = 0.2, 0.546 and 0.9 TeV~\cite{UA5} are reported, but a direct comparison with them is not straightforward due to different baryon number in the initial state. The E735 Collaboration provides measurements only for \pbar and not for p yields. Hence the proton-to-pion ratio is computed as 2\pbar/($\pi^{+}$+$\pi^{-}$). In addition, the E735 results for the proton-to-pion ratio are shown in Fig.~\ref{fig:RATIO:yield} only for \sqrts = 1.8 TeV because at the other energies the \pbar spectra include contributions of feed-down from weak decays and are not directly comparable with the measurements provided by the other experiments. 
For \sqrts $>$ 0.9 TeV, no dependence on the centre-of-mass energy of the (K$^{+}$+K$^{-}$)/($\pi^{+}$+$\pi^{-})$ and (p+\pbar)/($\pi^{+}$+$\pi^{-}$) ratios is observed within uncertainties.

In Fig.~\ref{fig:MEANPT:mass} the average transverse momenta \ptmean\xspace of pions, kaons, and protons, extracted from the sum of particle and antiparticle spectra, as a function of the centre-of-mass energy are reported.  Results at mid-rapidity in proton-proton collisions from  ALICE at \sqrts = 0.9, 2.76~\cite{900GeVspectra, 276ALICESpectra} and 7 TeV, CMS at \sqrts = 0.9, 2.76 and 7 TeV \cite{CMS_spectra} and PHENIX at \sqrts = 62.4 and 200 GeV~\cite{PHENIX_2010} are shown. In addition measurements obtained with E735 at \sqrts = 0.3, 0.54, 1 and 1.8 TeV~\cite{e735} and UA5 at \sqrts = 0.2, 0.546, 0.9 TeV \cite{UA5} in proton-antiproton collisions are also reported. The values of  \ptmean\xspace of \pbar from E735 are not shown since the spectra include contributions of feed-down from weak decays and hence are not directly comparable with the values provided by the other experiments. A slight increase of \ptmean\xspace with increasing  centre-of-mass energy is observed.
This rising trend is in particular apparent for \sqrts $>$ 0.9 TeV and it could be related to the increasing importance of hard processes at these energies.
At \sqrts = 7 TeV, the ALICE and CMS results are consistent within uncertainties except for the proton \ptmean.
This discrepancy is mostly due to the difference in the shape of the proton spectra seen in Fig.~\ref{fig:ALICE:CMS}, rather than to the extrapolation to the unmeasured \pt\xspace range: a 13\% difference is observed on the \ptmean\xspace values calculated from the ALICE and CMS data points in the  common \pt\xspace range.

\section{Comparison to models}
\begin{figure}
\centering

\includegraphics[trim=0cm 0.cm 0cm 0cm, clip=true, width=8.5cm]{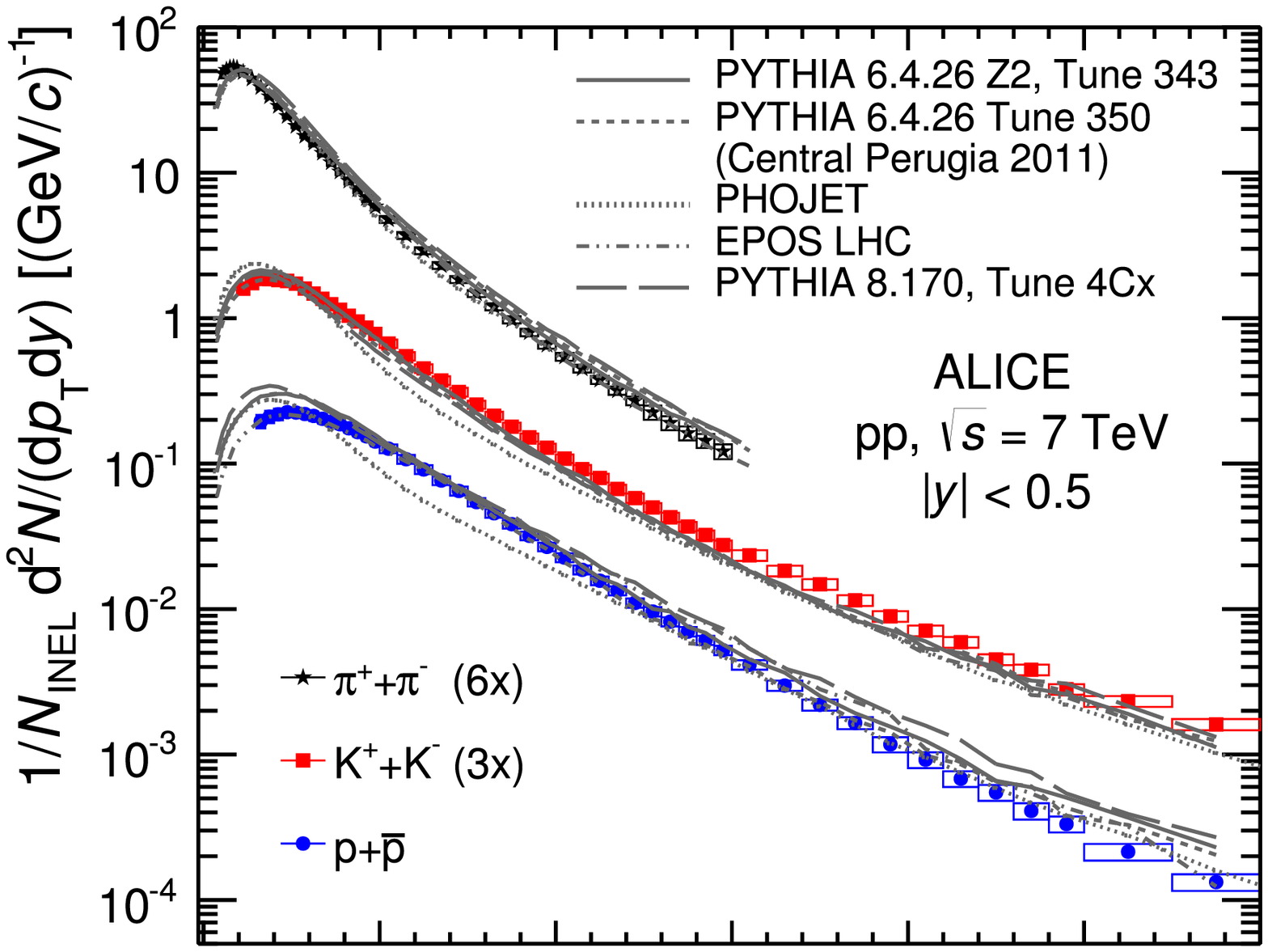} 
\includegraphics[trim=0cm 0cm 0cm 0.7cm, clip=true, width=8.5cm]{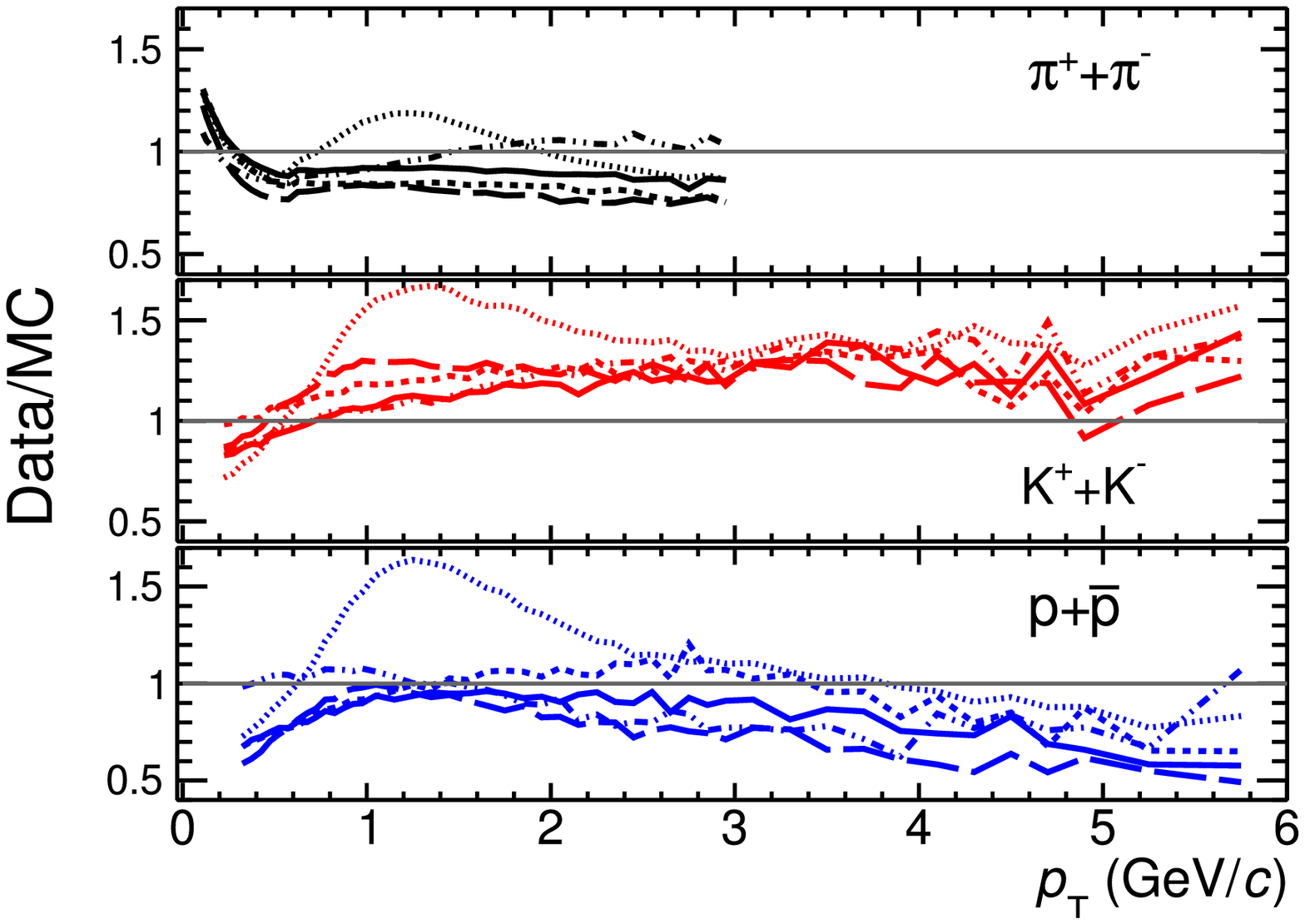}

\caption{Top panel: Measured \pt\xspace spectra of pions, kaons and protons, sum of particles and antiparticles, compared to PYTHIA6-Z2, PYTHIA6-CentralPerugia2011, PYTHIA8, EPOS LHC and PHOJET Monte-Carlo calculations. Statistical (vertical error bars) and systematic (open boxes) uncertainties are reported for the measured spectra. Bottom panels: ratios between data and Monte-Carlo calculations.}
\label{fig:COMB:pythia}
\end{figure}

 The comparison between the measured \pt\xspace spectra of $\pi$, K and p and the calculations of QCD-inspired Monte-Carlo event generators gives useful information on hadron production mechanisms. Figure~\ref{fig:COMB:pythia} shows the comparison of the measured pion, kaon and proton \pt\xspace spectra, sum of particles and antiparticles, with two tunes of the PYTHIA6 generator (PYTHIA6-CentralPerugia2011~\cite{PYTHIA-P2011} and PYTHIA6-Z2~\cite{PYTHIA-Z2})\footnote{The PYTHIA6 tunes are simulated using Rivet \cite{rivet}, a toolkit for validation of Monte-Carlo event generators.}, PYTHIA8 tune 4Cx~\cite{PYTHIA8,PYTHIA8i4Cx}, EPOS LHC~\cite{EPOS,EPOS1} and PHOJET~\cite{PHOJET}.\\ 
 These event generators are often used and tested to describe hadron collisions at high energies. PYTHIA is a general-purpose pQCD-based event generator, which uses a factorized perturbative expansion for the hardest parton-parton interaction, combined with parton showers and detailed models of hadronization and multiparton interactions. All presented PYTHIA tunes use a color reconnection mechanism~\cite{PYTHIA} which can mimic effects similar to that induced by collective flow in Pb-Pb collisions~\cite{CR}. In both PHOJET and EPOS, which are microscopic models that utilize the color-exchange mechanism of string excitation, the hadronic interactions are treated in terms of Reggeon and Pomeron exchanges.\\
PYTHIA6-Z2 tune is based on the first measurement of multiplicity distributions in minimum-bias pp collisions at \sqrts~=~900~GeV at the LHC. In the CentralPerugia2011 tuning both LEP fragmentation functions and minimun-bias charged particle multiplicity and underlying event data from the LHC are used. Both PYTHIA8 and EPOS LHC are tuned to reproduce the existing data available from the LHC (e.g. multiplicity and, for EPOS, also identified hadron production up to 1 \gevc for pions and kaons and up to 1.5 \gevc for protons). The PHOJET parameters are not retuned using the LHC data.\\
The measured pion \pt\xspace spectrum is reproduced by EPOS within 15\% over the whole \pt\xspace range. 
PYTHIA6-Z2, PYTHIA6-CentralPerugia2011 and PYTHIA8 show similar trends. They correctly predict the shapes of the pion spectra for \pt\xspace $>$ 500 MeV/c, overestimating the data by about 10\%, 20\% and 25\%, respectively, while the shapes differ from data for \pt\xspace $<$ 200 \mevc (the ratios are not flat) and the yields are underestimated by up to 30\%.
The PHOJET generator does not provide a satisfactory description of the measured spectrum shape for any of the particle species. The deviations from the data show a maximum for \pt\xspace$\sim$ 1.2\gevc and are more pronounced for kaons and protons than for pions.
All the tested Monte-Carlo generators underestimate the kaon yield by about 20-30\% for \pt\xspace $>$ 600 \mevc while for \pt\xspace$<$ 400 \mevc they overestimate the data by up to 30\%. A similar deviation is observed by the  ALICE collaboration also for other strange particle species  with a hierarchy depending on the strangeness content \cite{ALICE_s2}.
The proton yield is well described by EPOS only at low transverse momenta (\pt\xspace$<$ 1\gevc), while the generator tends to overestimate the data by up to 30\% at higher \pt.
None of the three PYTHIA tunes describes the shape of the proton spectrum in the full \pt\xspace range.
All of them give a reasonable description of the yield in the range 1 $< $ \pt\xspace$ <$ 2\gevc, but they overestimate the data at lower and higher \pt\xspace by up to 40\%.

\begin{figure}
\centering
\includegraphics[width=12 cm]{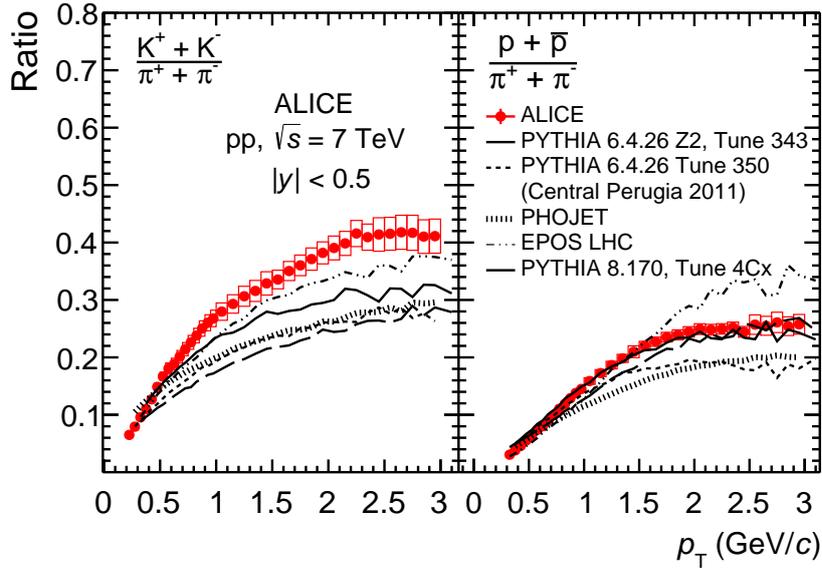}
\caption{ Measured (K$^{+}$+K$^{-}$)/($\pi^{+}$+$\pi^{-})$ (left) and (p+\pbar)/($\pi^{+}$+$\pi^{-})$ (right) ratios as a function of \pt \xspace compared to PYTHIA6-Z2, PYTHIA6-CentralPerugia2011, PYTHIA8, EPOS LHC and PHOJET calculations.  Statistical (vertical error bars) and systematic (open boxes) uncertainties are reported for the measured spectra.}
\label{fig:RATIO:pt}
\end{figure}

The comparison of the \pt-dependent particle ratios with models
allows the hadronization and soft parton interaction mechanisms implemented in the event generators to be tested.
In the left and right panels of Fig.~\ref{fig:RATIO:pt}, the measured (K$^{+}$+K$^{-}$)/($\pi^{+}$+$\pi^{-})$ and (p+\pbar)/($\pi^{+}$+$\pi^{-})$  ratios as a function of \pt\xspace  are compared with the same event generators shown in Fig.~\ref{fig:COMB:pythia}. 
The measured (K$^{+}$+K$^{-}$)/($\pi^{+}$+$\pi^{-})$ ratio increases from 0.05 at \pt\xspace= 0.2\gevc up  to 0.45 at \pt\xspace$\sim$ 3\gevc with a slope that decreases with increasing \pt. All the models underestimate the data at high momenta, with EPOS exhibiting the smallest deviation.
The measured (p+\pbar)/($\pi^{+}$+$\pi^{-})$  shows an increase from 0.03 at \pt\xspace = 0.3\gevc up to 0.25 at \pt\xspace$\sim$ 1.5\gevc while above this \pt\xspace it tends to flatten. The data are well described by PYTHIA6-Z2 while PYTHIA6-CentralPerugia2011, PHOJET and EPOS show a large deviation at high momenta. PYTHIA8 shows a smaller deviation over the whole momentum range even if, as seen in Fig.~\ref{fig:COMB:pythia}, it overestimates both pion and proton spectra.

The comparison between data and Monte-Carlo calculations shows that the tunes of the generators based only on few global observables, such as the integrated charged hadron multiplicity, allow only for a partial description of the data. The high-precision measurements of identified charged hadron  \pt\xspace spectra reported here, which cover a wide momentum range in the central rapidity region, give useful information for a fine tuning of the Monte-Carlo generators and a better understanding of soft particle production mechanisms at LHC energies.

\section{Summary}

A detailed analysis  of primary $\pi^{\pm}$, K$^{\pm}$, p and \pbar production in proton-proton collisions at \sqrts = 7 TeV with the ALICE detector has been performed. 
Particle identification is performed using several techniques namely the specific ionization energy loss measured in the ITS and TPC, the time of flight measured with the TOF detector, the Cherenkov radiation measured in the HMPID and the kink topology identification of the weak decays of charged kaons. The combination of these techniques allows for precision measurements of the \pt\xspace spectra over a wide momentum range: from 0.1 up to 3\gevc for pions, from 0.2 up to 6\gevc for kaons and from 0.3 up to 6\gevc for protons.
A comparison of the ALICE results with similar measurements performed by the PHENIX Collaboration at RHIC shows that the  \pt-integrated yields increase with collision energy for all the measured particle species. 
A slight increase of the \ptmean \xspace with \sqrts is also observed.
 This  rising trend that becomes apparent at \sqrts $>$ 0.9 TeV is established by the higher \sqrts LHC data. It could be related to the increasing importance of hard processes at these energies. 
The \pt-integrated K/$\pi$ and p/$\pi$ ratios extend the measurements available at lower collision energies from SPS, Sp{\pbar}S and RHIC experiments showing a saturation above \sqrts = 0.9 TeV.
Finally, the \pt\xspace spectra and particle ratios have been compared with the calculations of QCD-inspired  Monte-Carlo models namely PYTHIA6-Z2, PYTHIA6-CentralPerugia2011, PYTHIA8, EPOS LHC and PHOJET. Even though  the shapes of the spectra are fairly well reproduced by all models (except PHOJET that fails to describe the spectrum shape of all the three hadron species), none of them can describe simultaneously the measured yields of pions, kaons and protons.
These results can be used for a better understanding of the hadron production mechanisms in pp interactions at LHC energies and could further constrain the parameters of the models.

\newenvironment{acknowledgement}{\relax}{\relax}
\begin{acknowledgement}
\section*{Acknowledgements}
The ALICE Collaboration would like to thank all its engineers and technicians for their invaluable contributions to the construction of the experiment and the CERN accelerator teams for the outstanding performance of the LHC complex.
The ALICE Collaboration gratefully acknowledges the resources and support provided by all Grid centres and the Worldwide LHC Computing Grid (WLCG) collaboration.
The ALICE Collaboration acknowledges the following funding agencies for their support in building and
running the ALICE detector:
State Committee of Science,  World Federation of Scientists (WFS)
and Swiss Fonds Kidagan, Armenia,
Conselho Nacional de Desenvolvimento Cient\'{\i}fico e Tecnol\'{o}gico (CNPq), Financiadora de Estudos e Projetos (FINEP),
Funda\c{c}\~{a}o de Amparo \`{a} Pesquisa do Estado de S\~{a}o Paulo (FAPESP);
National Natural Science Foundation of China (NSFC), the Chinese Ministry of Education (CMOE)
and the Ministry of Science and Technology of China (MSTC);
Ministry of Education and Youth of the Czech Republic;
Danish Natural Science Research Council, the Carlsberg Foundation and the Danish National Research Foundation;
The European Research Council under the European Community's Seventh Framework Programme;
Helsinki Institute of Physics and the Academy of Finland;
French CNRS-IN2P3, the `Region Pays de Loire', `Region Alsace', `Region Auvergne' and CEA, France;
German Bundesministerium fur Bildung, Wissenschaft, Forschung und Technologie (BMBF) and the Helmholtz Association;
General Secretariat for Research and Technology, Ministry of
Development, Greece;
Hungarian Orszagos Tudomanyos Kutatasi Alappgrammok (OTKA) and National Office for Research and Technology (NKTH);
Department of Atomic Energy and Department of Science and Technology of the Government of India;
Istituto Nazionale di Fisica Nucleare (INFN) and Centro Fermi -
Museo Storico della Fisica e Centro Studi e Ricerche "Enrico
Fermi", Italy;
MEXT Grant-in-Aid for Specially Promoted Research, Ja\-pan;
Joint Institute for Nuclear Research, Dubna;
National Research Foundation of Korea (NRF);
Consejo Nacional de Cienca y Tecnologia (CONACYT), Direccion General de Asuntos del Personal Academico(DGAPA), M\'{e}xico, :Amerique Latine Formation academique – European Commission(ALFA-EC) and the EPLANET Program
(European Particle Physics Latin American Network)
Stichting voor Fundamenteel Onderzoek der Materie (FOM) and the Nederlandse Organisatie voor Wetenschappelijk Onderzoek (NWO), Netherlands;
Research Council of Norway (NFR);
National Science Centre, Poland;
Ministry of National Education/Institute for Atomic Physics and Consiliul Naţional al Cercetării Ştiinţifice - Executive Agency for Higher Education Research Development and Innovation Funding (CNCS-UEFISCDI) - Romania;
Ministry of Education and Science of Russian Federation, Russian
Academy of Sciences, Russian Federal Agency of Atomic Energy,
Russian Federal Agency for Science and Innovations and The Russian
Foundation for Basic Research;
Ministry of Education of Slovakia;
Department of Science and Technology, South Africa;
Centro de Investigaciones Energeticas, Medioambientales y Tecnologicas (CIEMAT), E-Infrastructure shared between Europe and Latin America (EELA), Ministerio de Econom\'{i}a y Competitividad (MINECO) of Spain, Xunta de Galicia (Conseller\'{\i}a de Educaci\'{o}n),
Centro de Aplicaciones Tecnológicas y Desarrollo Nuclear (CEA\-DEN), Cubaenerg\'{\i}a, Cuba, and IAEA (International Atomic Energy Agency);
Swedish Research Council (VR) and Knut $\&$ Alice Wallenberg
Foundation (KAW);
Ukraine Ministry of Education and Science;
United Kingdom Science and Technology Facilities Council (STFC);
The United States Department of Energy, the United States National
Science Foundation, the State of Texas, and the State of Ohio;
Ministry of Science, Education and Sports of Croatia and  Unity through Knowledge Fund, Croatia.
Council of Scientific and Industrial Research (CSIR), New Delhi, India
\end{acknowledgement}

\bibliographystyle{utphys}
\bibliography{mybib}

\newpage

\appendix
\section{The ALICE Collaboration}
\label{app:collab}



\begingroup
\small
\begin{flushleft}
J.~Adam\Irefn{org40}\And
D.~Adamov\'{a}\Irefn{org83}\And
M.M.~Aggarwal\Irefn{org87}\And
G.~Aglieri Rinella\Irefn{org36}\And
M.~Agnello\Irefn{org111}\And
N.~Agrawal\Irefn{org48}\And
Z.~Ahammed\Irefn{org131}\And
I.~Ahmed\Irefn{org16}\And
S.U.~Ahn\Irefn{org68}\And
I.~Aimo\Irefn{org94}\textsuperscript{,}\Irefn{org111}
% ...
\renewcommand\labelenumi{\textsuperscript{\theenumi}~}

\section*{Affiliation notes}
\renewcommand\theenumi{\roman{enumi}}
\begin{Authlist}
\item \Adef{0}Deceased
\item \Adef{idp5880112}{Also at: University of Kansas, Lawrence, Kansas, United States}
\end{Authlist}

\section*{Collaboration Institutes}
\renewcommand\theenumi{\arabic{enumi}~}
\begin{Authlist}

\item \Idef{org1}A.I. Alikhanyan National Science Laboratory (Yerevan Physics Institute) Foundation, Yerevan, Armenia
\item \Idef{org2}Benem\'{e}rita Universidad Aut\'{o}noma de Puebla, Puebla, Mexico
\item \Idef{org3}Bogolyubov Institute for Theoretical Physics, Kiev, Ukraine
\item \Idef{org4}Bose Institute, Department of Physics and Centre for Astroparticle Physics and Space Science (CAPSS), Kolkata, India
\item \Idef{org5}Budker Institute for Nuclear Physics, Novosibirsk, Russia
\item \Idef{org6}California Polytechnic State University, San Luis Obispo, California, United States
\item \Idef{org7}Central China Normal University, Wuhan, China
\item \Idef{org8}Centre de Calcul de l'IN2P3, Villeurbanne, France
\item \Idef{org9}Centro de Aplicaciones Tecnol\'{o}gicas y Desarrollo Nuclear (CEADEN), Havana, Cuba
\item \Idef{org10}Centro de Investigaciones Energ\'{e}ticas Medioambientales y Tecnol\'{o}gicas (CIEMAT), Madrid, Spain
\item \Idef{org11}Centro de Investigaci\'{o}n y de Estudios Avanzados (CINVESTAV), Mexico City and M\'{e}rida, Mexico
\item \Idef{org12}Centro Fermi - Museo Storico della Fisica e Centro Studi e Ricerche ``Enrico Fermi'', Rome, Italy
\item \Idef{org13}Chicago State University, Chicago, Illinois, USA
\item \Idef{org14}China Institute of Atomic Energy, Beijing, China
\item \Idef{org15}Commissariat \`{a} l'Energie Atomique, IRFU, Saclay, France
\item \Idef{org16}COMSATS Institute of Information Technology (CIIT), Islamabad, Pakistan
\item \Idef{org17}Departamento de F\'{\i}sica de Part\'{\i}culas and IGFAE, Universidad de Santiago de Compostela, Santiago de Compostela, Spain
\item \Idef{org18}Department of Physics and Technology, University of Bergen, Bergen, Norway
\item \Idef{org19}Department of Physics, Aligarh Muslim University, Aligarh, India
\item \Idef{org20}Department of Physics, Ohio State University, Columbus, Ohio, United States
\item \Idef{org21}Department of Physics, Sejong University, Seoul, South Korea
\item \Idef{org22}Department of Physics, University of Oslo, Oslo, Norway
\item \Idef{org23}Dipartimento di Elettrotecnica ed Elettronica del Politecnico, Bari, Italy
\item \Idef{org24}Dipartimento di Fisica dell'Universit\`{a} 'La Sapienza' and Sezione INFN Rome, Italy
\item \Idef{org25}Dipartimento di Fisica dell'Universit\`{a} and Sezione INFN, Cagliari, Italy
\item \Idef{org26}Dipartimento di Fisica dell'Universit\`{a} and Sezione INFN, Trieste, Italy
\item \Idef{org27}Dipartimento di Fisica dell'Universit\`{a} and Sezione INFN, Turin, Italy
\item \Idef{org28}Dipartimento di Fisica e Astronomia dell'Universit\`{a} and Sezione INFN, Bologna, Italy
\item \Idef{org29}Dipartimento di Fisica e Astronomia dell'Universit\`{a} and Sezione INFN, Catania, Italy
\item \Idef{org30}Dipartimento di Fisica e Astronomia dell'Universit\`{a} and Sezione INFN, Padova, Italy
\item \Idef{org31}Dipartimento di Fisica `E.R.~Caianiello' dell'Universit\`{a} and Gruppo Collegato INFN, Salerno, Italy
\item \Idef{org32}Dipartimento di Scienze e Innovazione Tecnologica dell'Universit\`{a} del  Piemonte Orientale and Gruppo Collegato INFN, Alessandria, Italy
\item \Idef{org33}Dipartimento Interateneo di Fisica `M.~Merlin' and Sezione INFN, Bari, Italy
\item \Idef{org34}Division of Experimental High Energy Physics, University of Lund, Lund, Sweden
\item \Idef{org35}Eberhard Karls Universit\"{a}t T\"{u}bingen, T\"{u}bingen, Germany
\item \Idef{org36}European Organization for Nuclear Research (CERN), Geneva, Switzerland
\item \Idef{org37}Excellence Cluster Universe, Technische Universit\"{a}t M\"{u}nchen, Munich, Germany
\item \Idef{org38}Faculty of Engineering, Bergen University College, Bergen, Norway
\item \Idef{org39}Faculty of Mathematics, Physics and Informatics, Comenius University, Bratislava, Slovakia
\item \Idef{org40}Faculty of Nuclear Sciences and Physical Engineering, Czech Technical University in Prague, Prague, Czech Republic
\item \Idef{org41}Faculty of Science, P.J.~\v{S}af\'{a}rik University, Ko\v{s}ice, Slovakia
\item \Idef{org42}Faculty of Technology, Buskerud and Vestfold University College, Vestfold, Norway
\item \Idef{org43}Frankfurt Institute for Advanced Studies, Johann Wolfgang Goethe-Universit\"{a}t Frankfurt, Frankfurt, Germany
\item \Idef{org44}Gangneung-Wonju National University, Gangneung, South Korea
\item \Idef{org45}Gauhati University, Department of Physics, Guwahati, India
\item \Idef{org46}Helsinki Institute of Physics (HIP), Helsinki, Finland
\item \Idef{org47}Hiroshima University, Hiroshima, Japan
\item \Idef{org48}Indian Institute of Technology Bombay (IIT), Mumbai, India
\item \Idef{org49}Indian Institute of Technology Indore, Indore (IITI), India
\item \Idef{org50}Inha University, Incheon, South Korea
\item \Idef{org51}Institut de Physique Nucl\'eaire d'Orsay (IPNO), Universit\'e Paris-Sud, CNRS-IN2P3, Orsay, France
\item \Idef{org52}Institut f\"{u}r Informatik, Johann Wolfgang Goethe-Universit\"{a}t Frankfurt, Frankfurt, Germany
\item \Idef{org53}Institut f\"{u}r Kernphysik, Johann Wolfgang Goethe-Universit\"{a}t Frankfurt, Frankfurt, Germany
\item \Idef{org54}Institut f\"{u}r Kernphysik, Westf\"{a}lische Wilhelms-Universit\"{a}t M\"{u}nster, M\"{u}nster, Germany
\item \Idef{org55}Institut Pluridisciplinaire Hubert Curien (IPHC), Universit\'{e} de Strasbourg, CNRS-IN2P3, Strasbourg, France
\item \Idef{org56}Institute for Nuclear Research, Academy of Sciences, Moscow, Russia
\item \Idef{org57}Institute for Subatomic Physics of Utrecht University, Utrecht, Netherlands
\item \Idef{org58}Institute for Theoretical and Experimental Physics, Moscow, Russia
\item \Idef{org59}Institute of Experimental Physics, Slovak Academy of Sciences, Ko\v{s}ice, Slovakia
\item \Idef{org60}Institute of Physics, Academy of Sciences of the Czech Republic, Prague, Czech Republic
\item \Idef{org61}Institute of Physics, Bhubaneswar, India
\item \Idef{org62}Institute of Space Science (ISS), Bucharest, Romania
\item \Idef{org63}Instituto de Ciencias Nucleares, Universidad Nacional Aut\'{o}noma de M\'{e}xico, Mexico City, Mexico
\item \Idef{org64}Instituto de F\'{\i}sica, Universidad Nacional Aut\'{o}noma de M\'{e}xico, Mexico City, Mexico
\item \Idef{org65}iThemba LABS, National Research Foundation, Somerset West, South Africa
\item \Idef{org66}Joint Institute for Nuclear Research (JINR), Dubna, Russia
\item \Idef{org67}Konkuk University, Seoul, South Korea
\item \Idef{org68}Korea Institute of Science and Technology Information, Daejeon, South Korea
\item \Idef{org69}KTO Karatay University, Konya, Turkey
\item \Idef{org70}Laboratoire de Physique Corpusculaire (LPC), Clermont Universit\'{e}, Universit\'{e} Blaise Pascal, CNRS--IN2P3, Clermont-Ferrand, France
\item \Idef{org71}Laboratoire de Physique Subatomique et de Cosmologie, Universit\'{e} Grenoble-Alpes, CNRS-IN2P3, Grenoble, France
\item \Idef{org72}Laboratori Nazionali di Frascati, INFN, Frascati, Italy
\item \Idef{org73}Laboratori Nazionali di Legnaro, INFN, Legnaro, Italy
\item \Idef{org74}Lawrence Berkeley National Laboratory, Berkeley, California, United States
\item \Idef{org75}Lawrence Livermore National Laboratory, Livermore, California, United States
\item \Idef{org76}Moscow Engineering Physics Institute, Moscow, Russia
\item \Idef{org77}National Centre for Nuclear Studies, Warsaw, Poland
\item \Idef{org78}National Institute for Physics and Nuclear Engineering, Bucharest, Romania
\item \Idef{org79}National Institute of Science Education and Research, Bhubaneswar, India
\item \Idef{org80}Niels Bohr Institute, University of Copenhagen, Copenhagen, Denmark
\item \Idef{org81}Nikhef, National Institute for Subatomic Physics, Amsterdam, Netherlands
\item \Idef{org82}Nuclear Physics Group, STFC Daresbury Laboratory, Daresbury, United Kingdom
\item \Idef{org83}Nuclear Physics Institute, Academy of Sciences of the Czech Republic, \v{R}e\v{z} u Prahy, Czech Republic
\item \Idef{org84}Oak Ridge National Laboratory, Oak Ridge, Tennessee, United States
\item \Idef{org85}Petersburg Nuclear Physics Institute, Gatchina, Russia
\item \Idef{org86}Physics Department, Creighton University, Omaha, Nebraska, United States
\item \Idef{org87}Physics Department, Panjab University, Chandigarh, India
\item \Idef{org88}Physics Department, University of Athens, Athens, Greece
\item \Idef{org89}Physics Department, University of Cape Town, Cape Town, South Africa
\item \Idef{org90}Physics Department, University of Jammu, Jammu, India
\item \Idef{org91}Physics Department, University of Rajasthan, Jaipur, India
\item \Idef{org92}Physik Department, Technische Universit\"{a}t M\"{u}nchen, Munich, Germany
\item \Idef{org93}Physikalisches Institut, Ruprecht-Karls-Universit\"{a}t Heidelberg, Heidelberg, Germany
\item \Idef{org94}Politecnico di Torino, Turin, Italy
\item \Idef{org95}Purdue University, West Lafayette, Indiana, United States
\item \Idef{org96}Pusan National University, Pusan, South Korea
\item \Idef{org97}Research Division and ExtreMe Matter Institute EMMI, GSI Helmholtzzentrum f\"ur Schwerionenforschung, Darmstadt, Germany
\item \Idef{org98}Rudjer Bo\v{s}kovi\'{c} Institute, Zagreb, Croatia
\item \Idef{org99}Russian Federal Nuclear Center (VNIIEF), Sarov, Russia
\item \Idef{org100}Russian Research Centre Kurchatov Institute, Moscow, Russia
\item \Idef{org101}Saha Institute of Nuclear Physics, Kolkata, India
\item \Idef{org102}School of Physics and Astronomy, University of Birmingham, Birmingham, United Kingdom
\item \Idef{org103}Secci\'{o}n F\'{\i}sica, Departamento de Ciencias, Pontificia Universidad Cat\'{o}lica del Per\'{u}, Lima, Peru
\item \Idef{org104}Sezione INFN, Bari, Italy
\item \Idef{org105}Sezione INFN, Bologna, Italy
\item \Idef{org106}Sezione INFN, Cagliari, Italy
\item \Idef{org107}Sezione INFN, Catania, Italy
\item \Idef{org108}Sezione INFN, Padova, Italy
\item \Idef{org109}Sezione INFN, Rome, Italy
\item \Idef{org110}Sezione INFN, Trieste, Italy
\item \Idef{org111}Sezione INFN, Turin, Italy
\item \Idef{org112}SSC IHEP of NRC Kurchatov institute, Protvino, Russia
\item \Idef{org113}SUBATECH, Ecole des Mines de Nantes, Universit\'{e} de Nantes, CNRS-IN2P3, Nantes, France
\item \Idef{org114}Suranaree University of Technology, Nakhon Ratchasima, Thailand
\item \Idef{org115}Technical University of Split FESB, Split, Croatia
\item \Idef{org116}The Henryk Niewodniczanski Institute of Nuclear Physics, Polish Academy of Sciences, Cracow, Poland
\item \Idef{org117}The University of Texas at Austin, Physics Department, Austin, Texas, USA
\item \Idef{org118}Universidad Aut\'{o}noma de Sinaloa, Culiac\'{a}n, Mexico
\item \Idef{org119}Universidade de S\~{a}o Paulo (USP), S\~{a}o Paulo, Brazil
\item \Idef{org120}Universidade Estadual de Campinas (UNICAMP), Campinas, Brazil
\item \Idef{org121}University of Houston, Houston, Texas, United States
\item \Idef{org122}University of Jyv\"{a}skyl\"{a}, Jyv\"{a}skyl\"{a}, Finland
\item \Idef{org123}University of Liverpool, Liverpool, United Kingdom
\item \Idef{org124}University of Tennessee, Knoxville, Tennessee, United States
\item \Idef{org125}University of the Witwatersrand, Johannesburg, South Africa
\item \Idef{org126}University of Tokyo, Tokyo, Japan
\item \Idef{org127}University of Tsukuba, Tsukuba, Japan
\item \Idef{org128}University of Zagreb, Zagreb, Croatia
\item \Idef{org129}Universit\'{e} de Lyon, Universit\'{e} Lyon 1, CNRS/IN2P3, IPN-Lyon, Villeurbanne, France
\item \Idef{org130}V.~Fock Institute for Physics, St. Petersburg State University, St. Petersburg, Russia
\item \Idef{org131}Variable Energy Cyclotron Centre, Kolkata, India
\item \Idef{org132}Vin\v{c}a Institute of Nuclear Sciences, Belgrade, Serbia
\item \Idef{org133}Warsaw University of Technology, Warsaw, Poland
\item \Idef{org134}Wayne State University, Detroit, Michigan, United States
\item \Idef{org135}Wigner Research Centre for Physics, Hungarian Academy of Sciences, Budapest, Hungary
\item \Idef{org136}Yale University, New Haven, Connecticut, United States
\item \Idef{org137}Yonsei University, Seoul, South Korea
\item \Idef{org138}Zentrum f\"{u}r Technologietransfer und Telekommunikation (ZTT), Fachhochschule Worms, Worms, Germany
\end{Authlist}
\endgroup


\end{document}